\documentclass{article}
\usepackage[utf8]{inputenc}
\usepackage{authblk}
\usepackage{setspace}
\usepackage[margin=1.25in]{geometry}
\usepackage{graphicx}
\usepackage{subcaption}
\usepackage{amsmath}
\usepackage{lineno}

\usepackage[style=nejm, 
citestyle=numeric-comp,
sorting=none]{biblatex}
\usepackage{xcolor}
\usepackage{enumerate}
\usepackage{enumitem}
\usepackage{tabularx}
\usepackage{comment}
\usepackage{algorithm}
\usepackage{algpseudocode}
\usepackage{amsmath,amssymb}
\usepackage{adjustbox}
\usepackage{booktabs}
\usepackage{float}
\usepackage{subcaption}
\usepackage{tikz}
\usetikzlibrary{arrows.meta,backgrounds}

\title{Positional choice and robust collective behavior in fish schools: biohybrid experiments and modeling}
\author[1]{Vahagn Grigoryan}
\author[2]{Donato Romano} 
\author[3]{Cesare Stefanini}
\author[1,2]{Giulia De Masi}

\affil[1]{Sorbonne University Abu Dhabi, Abu Dhabi, UAE}
\affil[2]{BioRobotics Institute, Sant’Anna School of Advanced Studies, Pisa, Italy}
\affil[3]{Division of Computing and Mathematical Science, MBZUAI, Abu Dhabi, UAE}
\affil[*]{Address correspondence to: vahagnderunc@gmail.com}

\date{}

\begin{document}

\maketitle

\begin{abstract}
Collective behavior of fish schools is usually modeled on the assumption that each individual follows specific rules of motion that depend on its position and velocity relative to its neighbors. Although these models reproduce many schooling patterns observed in nature, it remains unclear whether the assumed rules are realistic at the individual level. To address this question, we first analyzed a set of  experiments in which a live fish interacted with four moving robotic fish in a tank, and measured the time it spent in each position relative to the robots. We then simulated this experiment, replacing the fish with an agent, and assessed the extent to which classical models agree with the experimental observations. An extensive exploration of the parameter space showed that these models closely matched the experimental observations, with a very specific choice of parameters. However, they were highly sensitive to the parameter values: a minor perturbation caused them to fail completely. To resolve this, we propose a new model that incorporates an additional exploratory term characterizing the agent's positional preference at every moment. This model proved robust to small errors in the parameters while successfully replicating the experiments. Furthermore, when generalised to multiple fish, the model reproduced schooling behaviour and common schooling patterns, while also exhibiting exploratory behaviour at both the individual and the school level. This shows that social cohesion coexists with individual exploratory behaviour, and that realistic models of collective behaviour should account for both social interactions and this exploratory component. 
\end{abstract}


\section{Introduction}
Collective motion is among the most widespread behaviours in the animal kingdom, and
fish schooling is one of its clearest expressions. Its interest is twofold. Biologically,
aggregation confers antipredator and foraging advantages on the individual, and
understanding it is central to behavioural ecology. Methodologically, schools are a
canonical instance of the broader principle that coherent global structure can arise from
simple local interactions with no centralised controller \cite{waldrop1992complexity,
kelly2009out}. This principle has been carried into engineering as a design strategy for
distributed systems \cite{bonabeau1999swarm}, where the mechanisms of self-organisation
observed in animal groups are mapped onto artificial counterparts in optimisation, task
allocation and robotics. Modelling collective behaviour is therefore not only a means of
explaining what animals do, but also a route to designing systems that are adaptive and
resilient without central coordination.

The modelling strategy that follows from this principle is to prescribe the individual
rule and let the collective pattern emerge, rather than prescribing the shape of the group
directly. Since the seminal formulation of Reynolds \cite{reynolds1987flocks}, in which
repulsion, alignment and attraction are combined into a single behavioural rule, a large
number of models have been proposed along these lines. They differ in how the interaction
is specified, as physical-like forces \cite{hartono2024stochastic,
blomqvist2012mathematical, connor2023fish}, as concentric behavioural zones
\cite{couzin2002collective, aoki1982simulation, oboshi2003simulation}, as a fixed pattern maintained around a
leader \cite{gervasi2004coordination}, or as a response function inferred directly from
trajectory data \cite{bialek2012statistical, calovi2015collective}, and in the level of
description adopted, ranging from individual agents to kinetic and hydrodynamic limits
\cite{carrillo2009double, ha2008particle}. More recent work has extended the framework to
incorporate the surrounding fluid \cite{mao2025hybrid} and to extract the individual
control law experimentally through virtual reality \cite{li2025reverse}. These models
reproduce the main observed dynamics, including schooling, swarming, milling and
predator-avoidance patterns, and in this sense the collective level is well accounted for.

What remains open is the inverse question. Because these models are validated on
group-level observables (polarization, school diameter, inter-individual distances, the persistence of collective states) agreement at the collective level does not establish
that the assumed individual rule is the one the animal actually applies. Several distinct
rules can generate indistinguishable group patterns, and the individual contribution is
difficult to isolate experimentally, since the behaviour of a focal fish is continuously
perturbed by the conspecifics it is schooling with. The place an individual chooses to
occupy within the group, and how that choice depends on the external context, has
consequently been overlooked \cite{romano2021individual}. Addressing it requires an
experimental setting in which the focal individual can be observed while the rest of the
group is held fixed and unresponsive.



From an ethological perspective, fish schooling is not simply the spatial aggregation of
several individuals, but a dynamic form of social organisation produced by continuous
decisions about movement, spacing and orientation relative to nearby conspecifics. A
distinction is commonly made between \emph{shoaling}, in which fish remain together
because of social attraction, and \emph{schooling}, in which group members also coordinate
their swimming direction and speed \cite{pitcher1993functions, ward2016sociality}. The
occurrence of these behaviours across phylogenetically distant fish taxa and ecological
contexts suggests that collective organisation has repeatedly been favoured by natural
selection.

Group living can provide several fitness advantages. These include reducing individual
predation risk through dilution, enhanced vigilance and predator-confusion effects;
improving foraging efficiency; facilitating access to social information; and, under some
conditions, reducing the energetic costs of locomotion \cite{krause2002living,
ward2016sociality}. Experimental evidence further indicates that predators can exert direct
selection on the degree of coordination expressed by prey groups
\cite{ioannou2012predatory}. However, aggregation also entails costs, such as competition
for food, increased conspicuousness and exposure to inaccurate social information.
Schooling should therefore be interpreted as a flexible and context-dependent behavioural
strategy rather than as a fixed motor response.

Fish schools are also paradigmatic biological complex systems. Their coordinated
dynamics emerge from local interactions among individuals, without centralised control
\cite{couzin2009collective}. Through these interactions, information can propagate across
the group, allowing individuals to respond to environmental variation even when only a
subset of group members initially detects the relevant cue
\cite{sumpter2008information, berdahl2013emergent}. Collective organisation is therefore
both the outcome of individual decisions and a mechanism that modifies the sensory and
informational environment experienced by each individual.

At the same time, biological individuals are not interchangeable particles. Their
responses can vary according to internal state, previous experience, risk sensitivity,
social phenotype and environmental context. Persistent individual differences may
influence group cohesion, leadership, information transfer and collective performance
\cite{jolles2020heterogeneity}. Moreover, empirical studies have shown that interactions
among schooling fish can be anisotropic, nonlinear and dependent on neighbour
configuration, and that apparent alignment at the group level does not necessarily imply
that individuals explicitly match the orientation of nearby fish
\cite{katz2011inferring}. This biological variability makes it necessary to test whether
the local rules assumed in collective-motion models correspond to the behaviours actually
expressed by individual animals.

This work aims to identify the individual behaviours that give rise to the collective behaviour observed in biological fish schools. 
Here we isolate a single biological fish interacting with 4 robotic fish that cannot react to the fish movements \cite{romano2021individual}. In particular, we are interested in the behavior in the presence of a stimulus, specifically repulsive (predator). To the best of our knowledge, both the experimental set-up and the modeling remain unexplored in the literature. 

This paper addresses the following research questions: 
1) Is the single fish really applying the rules we assume for schooling, like in the Reynolds model?
2) If we run a simulation applying the behavior of the single agent to multiple agents, do we observe schooling or other behaviors?
3) What is the difference in the presence of stimuli (in particular a repulsive stimulus such as a predator)? 

\begin{figure}[H]
    \centering
    \includegraphics[width=\linewidth]{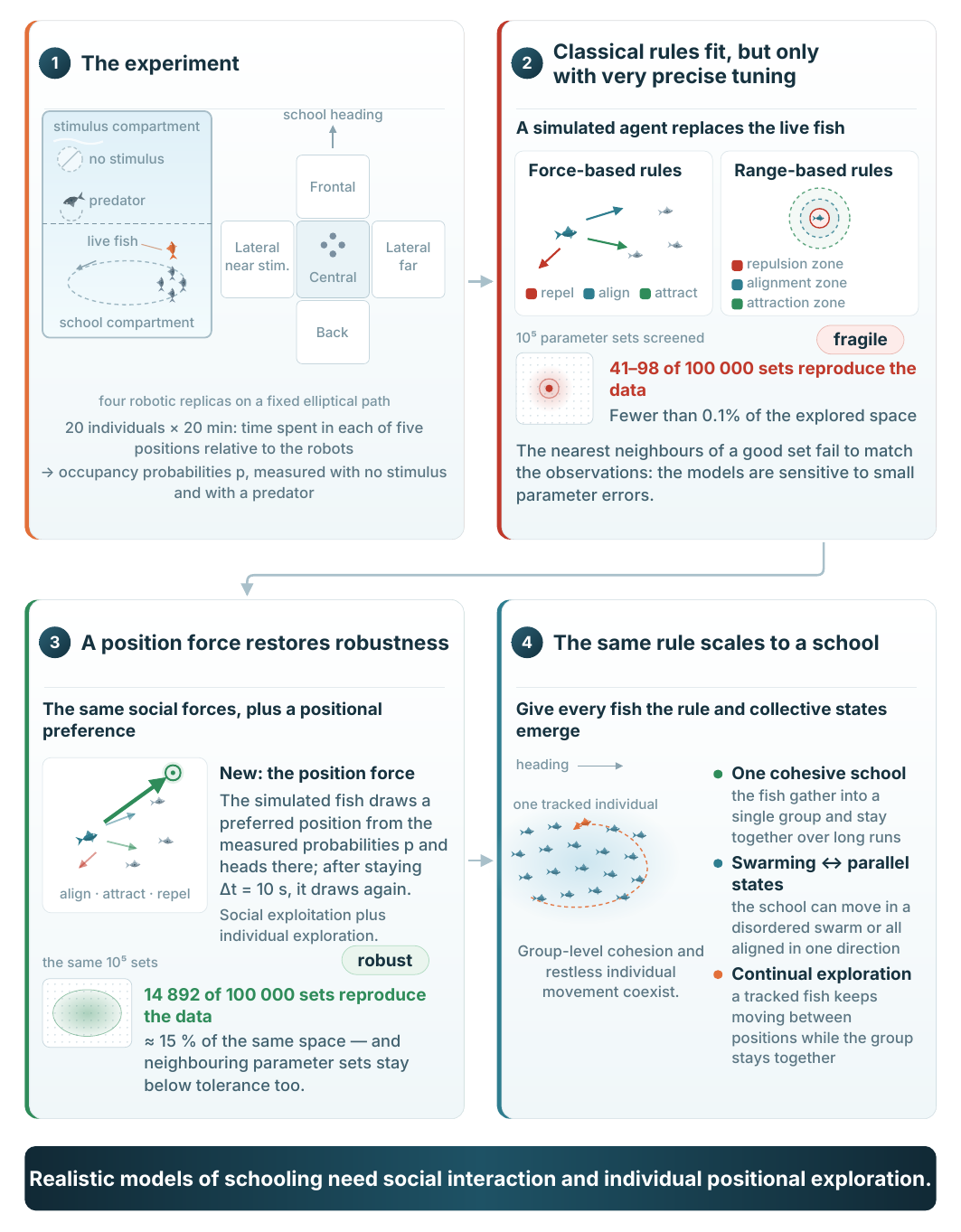} 
    \caption{Outline of the study}
    \label{visual-abstract_print}
\end{figure}

The remainder of the paper is organised as follows. Section~\ref{state-of-art} reviews existing models of collective motion, Section~\ref{materials-and-methods} describes the fish--robot experimental set-up, the models evaluated (two variants of a generalized force-based model, the range-based model, and the proposed generalized cohesion model), and the statistical procedures used to compare simulations with experiments: simulation-based parameter screening, local sensitivity analysis, and measures of school structure. Section~\ref{results} reports, for each model, the parameter combinations compatible with the experimental observations and their robustness to small parameter perturbations, in the no-stimuli context and, for the generalized force-based model, in the predator context; it then examines the behaviour of the generalized cohesion model in groups of multiple fish. Section~\ref{discussion} interprets these findings from modelling, ethological and evolutionary perspectives and outlines the limitations of the study. Finally, Section~\ref{conclusions} summarises the main conclusions. An overview of the study is given in Figure \ref{visual-abstract_print}.

\section{State of Art}
\label{state-of-art}

Recent efforts to model collective animal behavior, some also in predator-prey systems, have followed several complementary approaches. The seminal work on schooling behavior was introduced by Reynolds in 1987, with the behavioral rules typically formulated as follows \cite{reynolds1987flocks}: (1) Collision Avoidance (repulsion): avoid collisions with nearby individuals; (2) Velocity Matching (alignment): attempt to match the velocity of nearby individuals; (3) Flock Centering (attraction or flocking): attempt to remain close to nearby individuals. Such models also typically incorporate a stochastic component to account for variability in the behavior of individual agents.

A number of models employ a \textit{force-based framework} analogous to physical interactions. For example, the authors of \cite{hartono2024stochastic} proposed a differential equation model in which attraction, repulsion, and alignment forces are defined using weighted formulations, inspired by the Lennard-Jones potential and the law of gravitation. The model also incorporates a stochastic component. In addition, a repulsive predator force was introduced, and two predator hunting strategies were investigated. The model successfully reproduced natural avoidance patterns and enabled the estimation of survival probabilities.

In contrast, an earlier work \cite{uchitane2012ordinary} (later refined \cite{linh2015quantitative}) focused primarily on the theoretical analysis of the same model through the study and proof of the existence of local solutions and, under certain conditions, global solutions. The authors subsequently \cite{linh2015quantitative} examined how each parameter influences the geometric structure of the school, characterized in terms of: (1) distance to schoolmates, (2) minimum distance (the shortest distance between any two fish), (3) mean distance to schoolmates, (4) school diameter, (5) variance of fish velocities, and (6) the \(\epsilon\)-graph of the school, in which the vertices at time (t) are all the positions of fish and two vertices are connected if the distance between the corresponding fish is less than \(\epsilon\). The number of connected components of this graph represents the number of sub-schools. The study also investigated the effects of noise on the system. Incorporating both biological detail and environmental complexity, the authors of \cite{blomqvist2012mathematical} introduced a model comprising attraction, repulsion, and flocking forces acting on each individual. The attraction and repulsion forces associated with a single neighbour are respectively linear with distance and sigmoid-shaped, and are summed over all neighbours within a defined metric neighbourhood. The flocking force draws an individual toward the flock's center of mass and is linear with respect to the distance from the center. The model also incorporates multiple predators, with corresponding sigmoid-shaped avoidance forces acting on prey, as well as static obstacles in the environment. It is distinguished by its simulation of specific predator attack strategies and its assessment of the associated risks faced by prey.

Also grounded in a \textit{force-based framework}, another work \cite{connor2023fish} proposed the fish-inspired robotic algorithm (FIRA), which combines the classical attraction, alignment, and repulsion behaviors with additional predator avoidance, foraging,
general obstacle avoidance, and wandering forces, each formulated as a weighted,
distance-dependent term summed over an agent's field of view. Unlike prior force-based models, FIRA is explicitly designed to address the practical constraints of underwater swarm robotics: (1) an exponential attraction term and a multiplicative-inverse repulsion term bias agents toward stronger, more responsive
behavior at close range, (2) prediction terms compensate for sensor and processing
delay, (3) a short-term "fish memory" retains the last known position, velocity,
and detection time of neighbours, predators, food, and obstacles so that behavior
persists when perception is momentarily lost. A further distinguishing feature is
its reliance on indirect communication. FIRA was benchmarked in a Unity3D digital-twin simulation of a bio-inspired robotic fish against the Boids and artificial fish swarm algorithms across various types of trials. It achieved substantially fewer collisions and greater environment exploration than both baselines, and exhibited emergent milling and shoaling behaviors resembling real schooling fish.

A rigorous control-theoretic formalization of the \textit{force-based framework}
was provided in another work \cite{olfati2006flocking}, which recast Reynolds' three rules as a distributed feedback law for double-integrator agents and analyzed its convergence. Each agent's control input consists of three terms: a gradient-based term derived from a collective potential function, a velocity consensus (alignment) term acting as a damping force, and a navigational feedback term toward a group objective, represented by a virtual \(\gamma\)-agent. The obstacles are handled through a \textit{multi-species} construction, in which each physical agent (\(\alpha\)-agent) generates a virtual kinematic \(\beta\)-agent at its projection onto the boundary of a nearby obstacle, and interacts with it through
a potential of the same form; this allows obstacle avoidance to be treated with the same analytical machinery as inter-agent interaction. Three algorithms were proposed and analyzed. The first, comprising only the gradient and alignment terms, is shown to formally embody all three of Reynolds' rules, but is proven to lead to regular fragmentation. The second algorithm adds navigational feedback and is proven to keep the group cohesive, to guarantee asymptotic velocity matching among all agents and to remain collision-free under an explicit bound on the initial energy. The third algorithm incorporates the \(\beta\)-agent terms and achieves flocking in the presence of obstacles.

Some models adopt an approach of defining zones/ranges of the aforementioned forces (\textit{Range-based models}). In the article \cite{couzin2002collective}, each individual is characterized by repulsion, alignment, and attraction zones, with repulsion taking the highest priority. Thus, if at least one neighbor is located within the repulsion zone, only the repulsive force is applied. If the repulsion zone is empty but neighbors are present in either the alignment or attraction zones, only the corresponding force is applied. If neighbors are present in both zones, the resulting force is obtained by averaging the alignment and attraction forces. Neighbors within the alignment and attraction zones are considered only if they lie within the fish's field of perception. Additionally, individuals have a maximum turning rate. Based on the selected parameter values, four types of collective dynamical behavior were identified and characterized: \textit{Swarm}, \textit{Toroidal Milling}, \textit{Dynamic parallel group}, and \textit{Highly parallel group}. The study also investigated how changes in individual behavioral parameters during a simulation can induce transitions between different collective behavioral states. Furthermore, it analyzed how variations in parameter values among individuals within a group affect their spatial positions. Another agent-based model inspired by \textit{behavioral zones} is presented in the work \cite{oboshi2003simulation}, representing a major approach derived from Aoki's model \cite{aoki1982simulation}. In this framework, individual behavior is partitioned into zones of avoidance, parallel (analogous to alignment), and attraction. A key feature of the model is that each agent updates its direction based on a randomly selected neighbor rather than the entire group. Another major distinction from other models is that individual speeds are not fixed, but instead follow a gamma distribution, as proposed in \cite{aoki1980analysis}. This framework was employed to simulate predator-prey systems and reproduce realistic prey behavior.

An intriguing \textit{pattern-based} approach was proposed in the paper \cite{gervasi2004coordination}, in which agents (robots) follow a leader and seek to maintain a predefined formation around it. The formation pattern is dynamically adjusted at each time step according to both the leader's position and the barycenter of the group. The study investigated which formation patterns facilitate successful coordinated movement, providing practical insights into swarm robotics.

From a \textit{data-driven} perspective, the authors of \cite{bialek2012statistical} adopted an empirical approach by deriving a model of bird flocking from experimental observations. Rather than explicitly defining interaction forces, the authors inferred a probability distribution over the velocity directions of individuals and subsequently adjusted the model parameters to closely reproduce observed flocking behavior. Building on this data-driven tradition, the paper \cite{calovi2015collective} employed a model originally inferred from trajectory data in the work \cite{gautrais2012deciphering}. In this model, each individual's angular velocity evolves according to a stochastic differential equation driven by a response function that linearly combines alignment and attraction forces. These forces are weighted by corresponding coefficients and are computed over the first shell of Voronoi neighbors, rather than within a fixed metric or topological range. A key distinguishing feature of the model is an angular modulation term that weights the influence of each neighbor according to whether it is located ahead of or behind the focal fish. This breaks the action--reaction symmetry assumed in most force-based models and enables the model to reproduce not only schooling and swarming states, but also milling and elongated winding states. The authors used the model to investigate how the school responds to perturbations across different collective states, quantified its susceptibility to such perturbations, and examined its relationship with the intrinsic fluctuations within the school.

Some papers focus not on individual behavior, but rather on a macroscopical description of the system. For example, in the article \cite{carrillo2009double}, particles experience Rayleigh friction in addition to attraction/repulsion based on Morse potential, which is then used to derive a \textit{macroscopical description}. A further contribution \cite{ha2008particle} focused on the Cucker--Smale model (\cite{cucker2007mathematics}, \cite{cucker2007emergent}). In this model, each agent updates its velocity at every time step by adding a weighted average of the velocity differences between itself and its neighbors. The weights depend on the inter-agent distances and are chosen as non-increasing functions, commonly taking a form similar to an inverse-quadratic function. The study analyzed the model by examining fluctuations with respect to the center of the school and investigating their asymptotic behavior. A kinetic description of the model was subsequently introduced, and its time-asymptotic behavior was studied. This kinetic approach was then used to derive a hydrodynamic description of flocking.

A more recent and experimentally driven direction employs \textit{virtual reality (VR)-based reverse engineering} to directly extract the rule governing individual behavior \cite{li2025reverse}. The authors used an immersive volumetric VR system, in which a real zebrafish interacts with a virtual conspecific whose speed, heading, tail-beat timing, update rate, visibility, and spatial configuration
relative to the focal fish can be independently and precisely manipulated. By
networking VR arenas together, pairs of real fish were also able to interact within the same virtual world, allowing the authors to confirm that social responses in VR reproduce those observed in physical tanks. This experimental control enabled them to reverse-engineer a remarkably simple proportional-derivative control law, termed bioPD, that reproduces zebrafish social response with high fidelity. The authors conducted Turing-test trials and observed that real fish schooled with a bioPD-controlled virtual fish just as they did with another real fish. The model's generality was further established by (1)
benchmarking it against a state-of-the-art model predictive controller in a pursuit task across three robotic platforms (aerial, terrestrial, and aquatic), where bioPD achieved near-optimal performance, and (2) demonstrating that it scales across several juvenile fish species, group sizes, and multi-robot tasks.

Another recent work seeks to integrate the surrounding fluid dynamics directly into the
agent-based framework
\cite{mao2025hybrid}. It proposed a hybrid numerical model that couples a
self-propelled particle (SPP) model, built on the local-attraction framework
of the work \cite{strombom2011collective}, with a computational fluid dynamics (CFD)
solver. The SPP layer works much like earlier force-based models: each
individual's heading is updated from a weighted combination of directional
inertia, local attraction and alignment to nearby neighbours, local
repulsion, and attraction to the group centroid. The key difference is that
this SPP output is not used to move the agent directly. Instead, it sets a
target direction and speed for a simulated, flexible fish body, which the
CFD solver then drives through the fluid by resolving the actual
hydrodynamic forces acting on it. The resulting motion, shaped by the
surrounding flow rather than the SPP rules alone, is fed back into the SPP
model at the next time step, closing the loop between behavioural rules and
fluid forces. To keep this coupling physically realistic, the authors also
added a collision-avoidance mechanism and speed/turning-rate limits that
reflect the finite size and manoeuvrability of a real fish. Testing the model on groups of 30--50 individuals, the authors reproduced tight schooling, two sparse schooling configurations, and a milling pattern by varying the SPP alignment and attraction weights. Due to hydrodynamic interactions, both individual and group-average swimming speeds fluctuated substantially in every mode, in
contrast to the constant-speed assumption of traditional SPP models. The study also found that fish in all collective modes swam faster and at lower energetic cost than an equivalent solitary swimmer.

While the models surveyed above generally aim to reproduce cohesive collective
states such as schooling or milling, a related line of swarm robotics work exploits the same underlying force framework toward the opposite goal: dispersion rather than aggregation. This behavior, termed \textit{antiflocking}, was characterized in the article \cite{miao2010applying} as three heuristic rules: selfishness, de-centering, and collision avoidance, which mirror Reynolds' original repulsion, alignment, and attraction rules but tuned to keep agents apart rather than together. Building on this concept, the authors of \cite{xueobstacle} proposed a distributed antiflocking algorithm for multi-agent area coverage in environments with unknown, unmapped obstacles. Each agent leverages an adapted Bug-type navigation strategy that triggers whenever an obstacle is detected, allowing the agent to follow the obstacle's boundary until it can safely resume coverage. A key distinguishing feature is the Opposite Flood-fill Algorithm (inspired by \cite{olcay2022spontaneous}), which reconstructs obstacle geometry without any prior knowledge or seed point inside the obstacle. Agents further coordinate through three shared information maps fused through neighbor-to-neighbor communication so that each agent maintains a globally consistent picture of the environment from only local
sensing. Extensive simulations demonstrated that the proposed approach can effectively navigate unknown obstacles while substantially reducing overlap in non-coverable regions. Moreover, the approach exhibited strong scalability under sensing and communication constraints, supporting its applicability to large-scale distributed systems.

\newpage
\section{Materials and Methods}
\label{materials-and-methods}

\begin{figure}[H]
\centering
\singlespacing
\begin{tikzpicture}[x=1cm,y=1cm]
 
\tikzset{
  bx/.style={draw=black!65, line width=0.35pt, rounded corners=2pt,
             anchor=north west, align=left, inner sep=4pt, font=\scriptsize},
  gut/.style={rotate=90, anchor=center, font=\footnotesize\bfseries, text=black!80},
  flow/.style={-{Latex[length=2.2mm,width=1.7mm]}, line width=0.6pt, black!70},
  grp/.style={draw=black!45, dashed, dash pattern=on 2pt off 1.6pt,
              rounded corners=3pt, line width=0.4pt},
  gtag/.style={anchor=north west, font=\scriptsize\bfseries, text=black!75},
}
 
\definecolor{bandA}{RGB}{238,240,243} \definecolor{nodeA}{RGB}{218,224,231}
\definecolor{bandB}{RGB}{233,241,247} \definecolor{nodeB}{RGB}{206,225,239}
\definecolor{bandC}{RGB}{233,244,238} \definecolor{nodeC}{RGB}{205,231,216}
\definecolor{bandD}{RGB}{250,242,231} \definecolor{nodeD}{RGB}{245,224,197}
\definecolor{bandE}{RGB}{243,238,246} \definecolor{nodeE}{RGB}{226,215,235}
\definecolor{gutcol}{RGB}{205,209,214}
 
\begin{scope}[on background layer]
  \fill[bandA, rounded corners=3pt] (0,0.000) rectangle (15.20,-2.612);
  \fill[bandB, rounded corners=3pt] (0,-2.792) rectangle (15.20,-6.140);
  \fill[bandC, rounded corners=3pt] (0,-6.320) rectangle (15.20,-11.002);
  \fill[bandD, rounded corners=3pt] (0,-11.182) rectangle (15.20,-15.545);
  \fill[bandE, rounded corners=3pt] (0,-15.725) rectangle (15.20,-17.993);
  \fill[gutcol, rounded corners=3pt] (0,0.000) rectangle (1.10,-2.612);
  \fill[gutcol, rounded corners=3pt] (0,-2.792) rectangle (1.10,-6.140);
  \fill[gutcol, rounded corners=3pt] (0,-6.320) rectangle (1.10,-11.002);
  \fill[gutcol, rounded corners=3pt] (0,-11.182) rectangle (1.10,-15.545);
  \fill[gutcol, rounded corners=3pt] (0,-15.725) rectangle (1.10,-17.993);
\end{scope}
 
\node[gut] at (0.55,-1.306) {MATERIALS};
\node[gut] at (0.55,-4.466) {EXPERIMENT};
\node[gut] at (0.55,-8.661) {SIMULATION};
\node[gut] at (0.55,-13.364) {DATA ANALYSIS};
\node[gut] at (0.55,-16.859) {OUTPUTS};
 
\draw[grp] (1.350,-7.476) rectangle (11.050,-10.842);
\node[gtag] at (1.450,-7.506) {No-stimulus context};
\draw[grp] (11.350,-7.476) rectangle (15.200,-10.842);
\node[gtag] at (11.450,-7.506) {Predator context};
 
\node[bx, fill=nodeA, text width=4.1400cm, minimum height=2.312cm]
  at (1.3500,-0.140) {{\bfseries Experimental arena}\\[2pt]
Tank $600\times600\times400$\,mm, water depth $300$\,mm, divided by a removable
opaque partition into a school compartment and a stimulus compartment
($600\times300$\,mm each).};
\node[bx, fill=nodeA, text width=4.1400cm, minimum height=2.312cm]
  at (6.0639,-0.140) {{\bfseries Robotic school}\\[2pt]
Four 3D-printed \emph{P.~innesi} replicas, $90^{\circ}$ apart on a $32$\,mm
circle, on an elliptical path ($195\times80$\,mm semi-axes) at
$10$\,mm\,s$^{-1}$.};
\node[bx, fill=nodeA, text width=4.1400cm, minimum height=2.312cm]
  at (10.7778,-0.140) {{\bfseries Live subject and stimulus}\\[2pt]
One neon tetra (\emph{P.~innesi}) per trial. Predator: robotic
\emph{C.~johanna} replica on a $25$\,mm semicircular path.};
\node[bx, fill=nodeB, text width=4.1400cm, minimum height=1.966cm]
  at (1.3500,-2.932) {{\bfseries Treatments}\\[2pt]
Two conditions, $20$ trials each on $20$ different individuals: \emph{no
stimulus} (baseline) and \emph{predator}.};
\node[bx, fill=nodeB, text width=4.1400cm, minimum height=1.966cm]
  at (6.0639,-2.932) {{\bfseries Trial protocol}\\[2pt]
$30$\,min schooling with the robots, partition in place $\rightarrow$ partition
removed $\rightarrow$ $20$\,min recording.};
\node[bx, fill=nodeB, text width=4.1400cm, minimum height=1.966cm]
  at (10.7778,-2.932) {{\bfseries Measurement}\\[2pt]
Time spent in each of five positions relative to the robotic school: frontal,
back, central, lateral near / far from the stimulus.};
\node[bx, fill=nodeB, text width=13.5700cm, minimum height=0.883cm]
  at (1.3500,-5.097) {{\bfseries Experimental data set} (\cite{romano2021individual})\;:\;
$X_{\mathrm{obs}}\in\mathbb{R}^{20\times5}$, one row per trial, rows rescaled to
$1200$\,s $\Longrightarrow$ occupancy probabilities $p_P$ (Eq.~\ref{position_probabilities}).};
\node[bx, fill=nodeC, text width=13.5700cm, minimum height=0.816cm]
  at (1.3500,-6.460) {{\bfseries Simulated counterpart}\;:\; same arena and
robotic trajectory, the live fish replaced by one agent at constant speed $s$,
$\tau=0.1$\,s; $20$ independent runs per parameter set $\Longrightarrow$
$X_{\mathrm{sim}}\in\mathbb{R}^{20\times5}$.};
\node[bx, fill=nodeC, text width=1.9350cm, minimum height=2.876cm]
  at (1.5000,-7.816) {{\bfseries Force-based, constant alignment}\\[2pt]
Alignment $+$ Lennard--Jones $+$ noise (Eq.~\ref{Reynold_equation1}).};
\node[bx, fill=nodeC, text width=1.9350cm, minimum height=2.876cm]
  at (3.8943,-7.816) {{\bfseries Force-based, positive LJ alignment}\\[2pt]
As at left, with an LJ-type weight $w(r)$.};
\node[bx, fill=nodeC, text width=1.9350cm, minimum height=2.876cm]
  at (6.2885,-7.816) {{\bfseries Couzin range-based model}\\[2pt]
Zones $r_r$, $\Delta r_o$, $\Delta r_a$; perception field; turning rate.};
\node[bx, fill=nodeC, text width=1.9350cm, minimum height=2.876cm]
  at (8.6828,-7.816) {{\bfseries Generalized cohesion (proposed)}\\[2pt]
Adds a position force $c_p p_i(t)$ toward a position drawn from $p_P$
(Eq.~\ref{mathematical_model_equation}).};
\node[bx, fill=nodeC, text width=3.2700cm, minimum height=2.876cm]
  at (11.5000,-7.816) {{\bfseries Force-based, constant alignment $+$ predator
avoidance}\\[2pt]
Repulsive sigmoid stimulus force with weight $c_P$, midpoint $r_0$, steepness
$\beta$ (Eq.~\ref{force-based-predator-eq}).};
\node[bx, fill=nodeD, text width=13.5700cm, minimum height=0.865cm]
  at (1.3500,-11.322) {{\bfseries Parameter screening (ABC-inspired)}\;:\;
$N=10^{5}$ parameter combinations per model, drawn uniformly or log-uniformly
over the ranges of Tables~\ref{parameters_lopez}--\ref{parameters_mathematical}, each simulated $20$ times.};
\node[bx, fill=nodeD, text width=6.4700cm, minimum height=1.685cm]
  at (1.3500,-12.387) {{\bfseries Discrepancy measure}\\[2pt]
$\rho_j=\frac{1}{400}\sum_{i=1}^{20}\sum_{i'=1}^{20}(X_{ij}-X'_{i'j})^2$,
\; $\rho=\frac{1}{5}\sum_{j=1}^{5}\rho_j$: every row compared with every row, so
$\rho$ compares distributions, not matched trials.};
\node[bx, fill=nodeD, text width=6.4700cm, minimum height=1.685cm]
  at (8.4478,-12.387) {{\bfseries Acceptance criterion}\\[2pt]
$\rho(X_{\mathrm{sim}},X_{\mathrm{obs}})<\delta$, with $\delta=10000$ (no
stimulus; self-error $7652.24$) and $\delta=8000$ (predator; $4762.57$). Runs
with zero frontal position are not analysed further.};
\node[bx, fill=nodeD, text width=13.5700cm, minimum height=1.112cm]
  at (1.3500,-14.273) {{\bfseries Local sensitivity analysis}\;:\; parameters
mapped onto $[0,1]^m$ (linearly, or on a $\log_{10}$ scale where sampling was
log-uniform); the five nearest neighbours of a well-performing combination, in
Euclidean distance $D$, are compared on $\rho$ and on the frontal-position mean.};
\node[bx, fill=nodeE, text width=6.4700cm, minimum height=1.968cm]
  at (1.3500,-15.865) {{\bfseries Robustness across models}\\[2pt]
Distribution of $\rho$, and combinations accepted out of $10^{5}$ (of which with
non-zero frontal position): force-based constant $62$ ($9$), positive LJ $41$
($4$), Couzin $98$ ($27$), generalized cohesion $14892$ ($14588$); predator
$182$.};
\node[bx, fill=nodeE, text width=6.4700cm, minimum height=1.968cm]
  at (8.4478,-15.865) {{\bfseries Generalization to a school of live fish}\\[2pt]
Generalized cohesion model with $N$ agents and neighbourhood radius $R$:
polarization (mean, max), school length and width against $N$, and current
versus desired position of a tracked individual.};
 
\draw[flow] (8.275,-2.612) -- (8.275,-2.792);
\draw[flow] (8.275,-4.897) -- (8.275,-5.097);
\draw[flow] (8.275,-6.140) -- (8.275,-6.320);
\draw[flow] (8.275,-7.276) -- (8.275,-7.476);
\draw[flow] (8.275,-11.002) -- (8.275,-11.182);
\draw[flow] (8.275,-12.187) -- (8.275,-12.387);
\draw[flow] (8.275,-14.073) -- (8.275,-14.273);
\draw[flow] (8.275,-15.545) -- (8.275,-15.725);
 
\end{tikzpicture}
\caption{Flowchart of experiment and model design.}
\label{fig:experimental_design}
\end{figure}
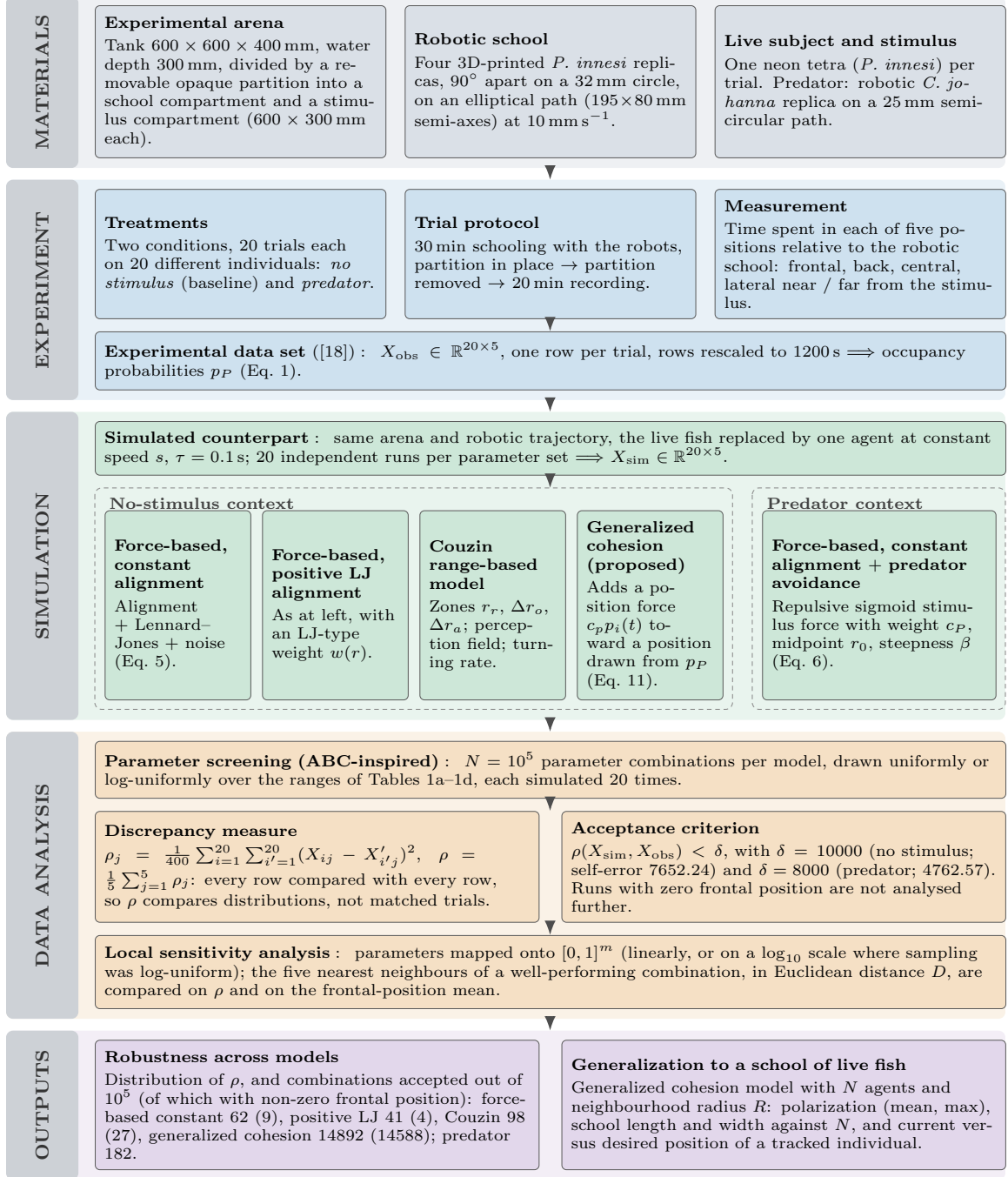

The procedure of the methodology is illustrated in Figure \ref{fig:experimental_design}.

\subsection{Experimental Design}

\subsubsection{Experimental set-up}
\label{SetUp}

The paper \cite{romano2021individual} investigated where individual neon tetras (\textit{Paracheirodon innesi}) choose to position themselves within a fish school under different ecological conditions, utilizing a robotic school of fish.

\begin{figure}[H]
    \centering
    \begin{subfigure}{0.48\textwidth}
        \includegraphics[width=\linewidth]{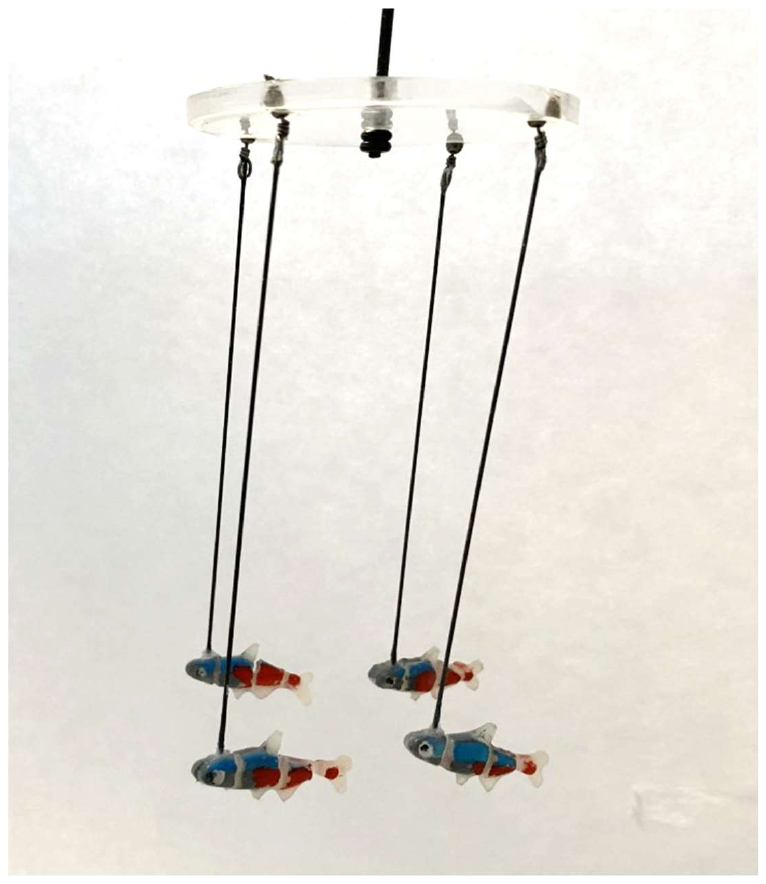}
        \caption{Four robotic fish.}
        \label{Robotic fishes}
    \end{subfigure}
    \begin{subfigure}{0.48\textwidth}
        \includegraphics[width=\linewidth]{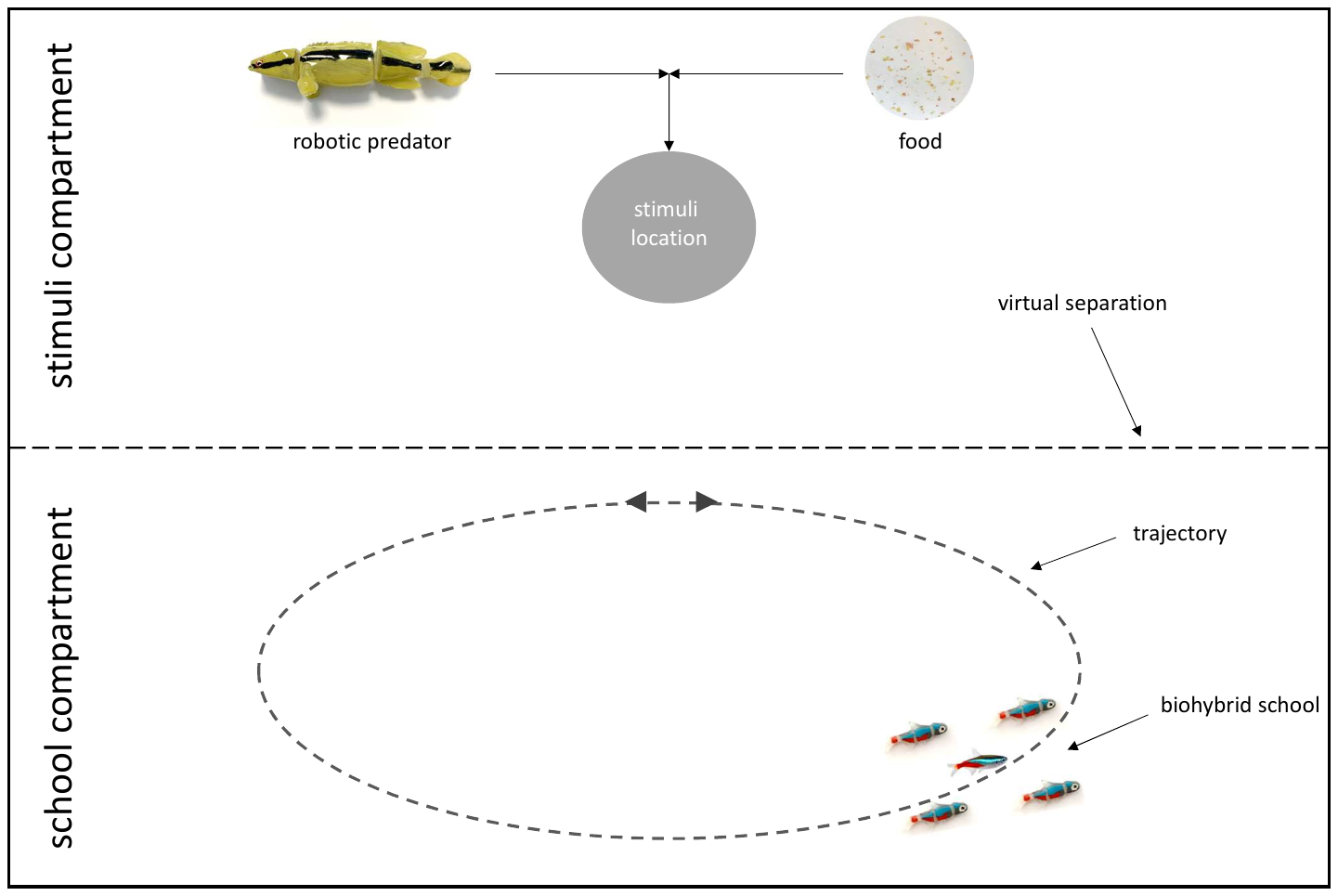}
        \caption{Trajectory and different stimuli.}
        \label{robotic fishes_trajectory}
    \end{subfigure}
    \caption{Experiment with robotic fish}
\end{figure}

Four fish replicas were 3D-printed. These four replicas were suspended from rods connected to a rotating overhead dish, arranged 90° apart on a 32mm-radius circle. The robotic school followed an elliptical trajectory (195mm × 80mm semi-axes, 10mm/s) inside the tank (Figure \ref{Robotic fishes}). 

A 600×600×400mm tank (water depth 300mm) was split by a removable opaque partition into a "school compartment," (600×300) where the test fish swam alongside the moving robotic school, and a "stimuli compartment," (600×300) where either no stimuli or the robotic predator was placed.

The predator stimulus was a robotic replica of \textit{Crenicichla johanna}, a natural predator of \textit{P. innesi}. A robotic arm moved the replica along a semicircular trajectory of 25 mm radius. It was placed centrally in the stimulus compartment and presented by removing the partition separating it from the school compartment; the no-stimulus condition followed the same procedure without introducing a cue, serving as baseline (Figure \ref{robotic fishes_trajectory}). 

This arrangement created five positions a real fish could occupy relative to the robot school: frontal, back, central, lateral (near the stimulus), and lateral (far from the stimulus) (Figure \ref{robotic_positions_definitions}).

For each condition (no external stimuli, and repulsive source (predator)), 20 experiments have been repeated with 20 different individuals. Each trial consisted of an initial 30-minute period with the opaque partition in place, during which the live fish schooled with the robots in the absence of stimuli. The partition was then removed to enable the presentation of stimuli, and over the subsequent 20 minutes, the researchers recorded the amount of time the fish spent in each of the five positions.

Based on the measured times spent in the specific positions, we here infer a probability for the fish to occupy one of these positions. 

\begin{equation}
    p_P = \frac{\sum_{i=1}^{20}T_{i, P}}{\sum_{j=1}^5 \sum_{i=1}^{20}T_{i, j}} 
    \label{position_probabilities}
\end{equation}

We simulated the same experiments replacing the live fish with an agent, and investigated several models the agent can follow.

\subsection{Models}

\subsubsection{Generalized force-based model} 

\paragraph{The Lennard-Jones Function}

Schooling behavior is commonly simulated using a particle interaction framework, analogous to the interactions between molecules or atoms. The interaction between two molecules or atoms is typically described by the Lennard-Jones potential function, which is also usually used in schooling behavior simulations. It is given by:

\[
V(r) = 4\epsilon \left[ \left( \frac{\sigma}{r} \right)^{12} - \left( \frac{\sigma}{r} \right)^{6} \right]
\]

where:
\begin{itemize}
    \item \( r \) is the distance between particles,
    \item \( \epsilon \) is the depth of the potential well,
    \item \( \sigma \) is the finite distance at which the inter-particle potential is zero.
\end{itemize}

The formula is often generalized by replacing the fixed exponents with a \textit{steepness} parameter, \(\alpha\):

\[
V(r) = 4\epsilon \left[ \left( \frac{\sigma}{r} \right)^{2\alpha} - \left( \frac{\sigma}{r} \right)^{\alpha} \right]
\]

By differentiating the potential function, we obtain the corresponding interaction force:

\begin{equation}
    F(r)=-\frac{dV}{dr}=-8\alpha\epsilon \frac{\sigma^{2\alpha}}{r^{2\alpha+1}}+4\alpha\epsilon \frac{\sigma^\alpha}{r^{\alpha+1}}
\label{Lennard_Jones_General}
\end{equation}

\paragraph{Simulating the generalized force-based model}

A large variety of models employ a force-based frame framework. One possible generalization of these models can be characterized as follows \cite{lopez2012behavioural}:
An individual fish moves at a constant speed \(s \text{ mm}/\text{s}\) and updates its direction based on its neighbors' influence and a stochastic noise term. The neighbour's have two types of influence: 1) alignment force: fish tries to align himself with the direction of its neighbours 2) repulsion/attraction forces: fish tries to get closer to its neighbors, meanwhile avoiding collisions with them.
The updates can be described by the following equations (Equation \ref{force-based-eq-0}, \ref{force-based-eq-0-update}).

\begin{equation}
    \mathbf{d}_i(t+\tau) = \frac{1}{n(t)} \sum_{j \in N_i(t)} w(|\mathbf{r}_{ij}(t)|) \mathbf{d}_j(t)
    + \frac{1}{n(t)} \sum_{j \in N_i(t)} f(|\mathbf{r}_{ij}(t)|) \frac{\mathbf{r}_{ij}(t)}{|\mathbf{r}_{ij}(t)|} + \eta_i(t).
    \label{force-based-eq-0}
\end{equation}

where:

\begin{itemize}
    \item \(\tau\) is a discrete time step (sec).
    \item \( N_i(t) \) is a defined neighborhood around \(i^{\text{th}}\) fish at the time \(t\), with \(n(t)\) neighbors.
    \item \(\mathbf{d}_i(t)\) is a directional unit vector of the \(i^{\text{th}}\) fish at the time \(t\),
    \item \(\mathbf{r}_{ij}(t) = \mathbf{r}_j(t)-\mathbf{r}_i(t)\) is the relative position of the \(j^{\text{th}}\) fish with respect to \(i^{\text{th}}\) fish.
    \item \(w(r)\) is a function defining relative range and weight of alignment.
    \item \(f(r)\) is a function defining relative range and weight of attraction/repulsion. 
    \item \(\eta_i(t)\) is a stochastic component. In all the equations here, it is added after the sum of the forces is normalized.
\end{itemize}

The preferred direction \(\mathbf{d}_i(t+\tau)\) is then normalized
to a unit vector.
For finding the new position:

\begin{equation}
    \mathbf{r}_i(t+\tau) = \mathbf{r}_i(t) + s\tau \mathbf{d}_i(t+\tau).
    \label{force-based-eq-0-update}
\end{equation}

We here divide time into discrete \(\tau=0.1 \text{ s}\) steps, to make the movement smoother.
We also consider representing the weighting constants (factors) separately from the functions \(w(r)\) and \(f(r)\), resulting in the following formulation:

\begin{equation}
    \mathbf{d}_i(t+\tau) = c_w\frac{1}{n(t)} \sum_{j \in N_i(t)} w(|\mathbf{r}_{ij}(t)|) \mathbf{d}_j(t)
    + c_f\frac{1}{n(t)} \sum_{j \in N_i(t)} f(|\mathbf{r}_{ij}(t)|) \frac{\mathbf{r}_{ij}(t)}{|\mathbf{r}_{ij}(t)|} + \eta_i(t)
    \label{Reynold_equation1}
\end{equation}

where \(c_w\) and \(c_f\) are constants.

\begin{enumerate}[label=\alph*.]
    \item The neighborhood \(N_i\) is defined to be the four robotic fish.
    \item For the function \(w(r)\), we use either a constant function, or a function similar to Lennard-Jones:
    \begin{enumerate}[label=\roman*.]
        \item \[
        w(r) = 1       
        \]
        \item \[
        w(r) = 8\alpha\epsilon \frac{\sigma^{2\alpha}}{r^{2\alpha+1}}+4\alpha\epsilon \frac{\sigma^\alpha}{r^{\alpha+1}} \quad \text{\cite{hartono2024stochastic}}
        \]
    \end{enumerate}
    
    \item For the function \(f(r)\), we can use the Lennard-Jones formula  ~(\ref{Lennard_Jones_General}). \cite{ferrante2014self}, \cite{amorim2021self}
    \item For the stochastic component \( \eta_i(t) \), we can use a Gaussian distribution.
    So \(\eta_i(t)\) is the vector \((\mathcal{N}_1(t), \mathcal{N}_2(t)\)) and both of its components are Gaussian with mean \(0\) and standard deviation \(\mu\).
\end{enumerate}

\begin{table}[H]
    \centering
    \caption{Ranges of parameters investigated, for every model.}
    \begin{subtable}[l]{0.98\textwidth}
    \centering
    \footnotesize
    \begin{tabular}{|l||l|l|}
        \hline
        Parameter & Description & Values \\ \hline
        \(s\) & Speed of the fish (mm per second) & \([5, 20]\) \\
        \(\epsilon\) &  Depth of the potential well of Lennard-Jones function & \(1\)\\
        \(\sigma\) & The finite distance at which the inter-particle potential is zero & \([1, 20]\) \\
        \(\alpha\) & Steepness of the Lennard-Jones function & \([1, 6]\) \\
        \(c_w\) & Alignment weight & \(1\)\\
        \(c_f\) & Attraction/repulsion weight & \(10^{[-2, 4]}\)\\
        \(\mu\) & Noise coefficient & \( [0, 1]\) \\
        \(\tau\) & Time step (seconds) & \(0.1\) \\
        \hline
    \end{tabular}
    \caption{\footnotesize Generalized force-based model, no-stimuli context, constant/positive LJ alignment}
    \label{parameters_lopez}
    \end{subtable}

    \vspace{0.5cm}
    
    \begin{subtable}[l]{0.98\textwidth}
    \centering
    \footnotesize
    \begin{tabularx}{\textwidth}{|l||X|l|}
        \hline
        Parameter & Description & Values \\ \hline
        \(s\) & Speed of the fish (mm per second) & \([5, 20]\) \\
        \(\epsilon\) &  Depth of the potential well of Lennard-Jones function & \(1\)\\
        \(\sigma\) & The finite distance at which the inter-particle potential is zero & \([1, 20]\) \\
        \(\alpha\) & Steepness of the Lennard-Jones function & \([1, 6]\) \\
        \(c_w\) & Alignment weight & \(1\)\\
        \(c_f\) & Attraction/repulsion weight & \(10^{[-2, 4]}\)\\
        \(\mu\) & Noise coefficient & \( [0, 1]\) \\
        \(\tau\) & Time step (seconds) & \(0.1\) \\
        \(c_P\) & Predator force weight & \(10^{[-2, 4]}\) \\
        \(r_0\) & Predator interaction midpoint (mm) & \([150, 450]\) \\
        \(\beta\) & Steepness of the sigmoid function & \([0.005, 0.1]\) \\
        \hline
    \end{tabularx}
    \caption{\footnotesize Generalized force-based model, predator context, constant alignment}
    \label{parameters_lopez_predator}
    \end{subtable}
    
    \vspace{0.5cm}
    
    \begin{subtable}[l]{0.98\textwidth}
    \centering
    \footnotesize
    \begin{tabularx}{\textwidth}{|l||X|l|}
        \hline
        Parameter & Description & Values \\ \hline
        \(r_r\) & Zone of repulsion & \([1, 25]\) \\
        \(\Delta r_o \, (r_o-r_r)\) &  Zone of alignment & \([0, 90]\)\\
        \(\Delta r_a \, (r_a-r_o)\) & Zone of attraction & \([0, 90]\) \\
        \(\alpha\) & Field of perception & \([200, 360]\) \\ 
        \(\theta\) & Turning rate & \([10, 180]\) \\
        \(s\) & Speed (unit per second) & \([5, 20]\)\\
        \(\sigma\) & Noise sd (rad) & \([0, 0.2]\)\\
        \(\tau\) & Time step (seconds) & \(0.1\) \\ \hline
    \end{tabularx}
    \caption{\footnotesize Range-based model, no-stimuli context.}
    \label{parameters_couzin}
    \end{subtable}
    
    \vspace{0.5cm}
    
    \begin{subtable}[l]{0.98\textwidth}
    \centering
    \footnotesize
    \begin{tabularx}{\textwidth}{|l||X| l|}
        \hline
        Parameter & Description & Values \\ \hline
        \(s\) & Speed of the fish (mm per second) & \([5, 20]\) \\
        \(\Delta t\) & Update delay of the desired position of the fish & \(10\) sec \\
        \(\epsilon\) &  Depth of the potential well of Lennard-Jones function & \(1\)\\
        \(\sigma\) & The finite distance at which the inter-particle potential is zero & \([1, 20]\) \\
        \(\alpha\) & Steepness of the Lennard-Jones function & \([1, 6]\) \\
        \(c_w\) & Alignment weight & \(1\)\\
        \(c_f\) & Attraction/repulsion weight & \(10^{[-2, 4]}\)\\
        \(c_p\) & Position force weight & \(10^{[-3,0]}\)\\
        \(\mu\) & Noise coefficient & \( [0, 1]\) \\
        \(\tau\) & Time step (seconds) & \(0.1\) \\
        \hline
    \end{tabularx}
    \caption{\footnotesize Generalized cohesion model, no-stimuli context, constant alignment.}
    \label{parameters_mathematical}
    \end{subtable}
\end{table}

The choice of parameters (Table \ref{parameters_lopez}) is determined by following criteria:

\begin{itemize}
\item \(s\): The robots move at a constant speed of \(10\text{mm}/\text{s}\); We consider it reasonable to investigate values from half this speed up to twice this speed.

\item \(\epsilon\): This parameter acts as a factor in the Lennard--Jones formula. We fix it to be \(1\) and vary the \(c_f\) parameter.

\item \(\sigma\): The quantity \(\sigma \cdot 2^{\frac{1}{\alpha}}\) represents the distance at which the attraction/repulsion force is zero, with repulsive behavior occurring at shorter distances. The distance between two robots is \(32\sqrt{2} \approx 45.25\). Therefore, for the agent to be able to pass between them, we require
\(
\sigma \cdot 2^{\frac{1}{\alpha}} < 16\sqrt{2}
\).

\item \(\alpha\): We consider the values till the original \(\alpha=6\).

\item \(c_w, c_f\): Since all robots move in the same direction, the alignment force is consistently high, (with a magnitude of \(1\) in case of constant alignment), whereas the attraction/repulsion force, particularly when acting as an attraction, is generally much smaller. Since the resulting vector is normalized at the end, only the ratio of the weights affects the resulting direction. Therefore, we fix \(c_w=1\) and vary \(c_f\).

\item \(\mu\): The noise is added to a normalized unit vector. Therefore, choosing a standard deviation greater than \(1\) would result in excessive noise, which was also verified empirically.
\end{itemize}

\paragraph{Adding a predator}

To model the predator escaping behavior, commonly a predator avoidance force is introduced. The directional unit vector is described by Equation \ref{force-based-predator-eq}.

\begin{equation}
\begin{aligned}
\mathbf{d}_i(t+\tau) &= c_w\frac{1}{n(t)} \sum_{j \in N_i(t)} w(|\mathbf{r}_{ij}(t)|) \mathbf{d}_j(t) + c_f\frac{1}{n(t)} \sum_{j \in N_i(t)} f(|\mathbf{r}_{ij}(t)|) \frac{\mathbf{r}_{ij}(t)}{|\mathbf{r}_{ij}(t)|} \\
&\quad + c_P g(|P_i(t) - r_i(t)|) \frac{P_i(t) - r_i(t)}{|P_i(t) - r_i(t)|} + \eta_i(t)
\end{aligned}
\label{force-based-predator-eq}
\end{equation}

And we choose,

\begin{equation*}
g(r) = - \frac{1}{1+e^{\beta (r-r_0)}} \quad \text{\cite{blomqvist2012mathematical}}
\end{equation*}

where,

\begin{itemize}
    \item \(P_i(t)\) is the position of the predator
    \item \(c_P\) is the weight of the introduced repulsive force
    \item \(\beta\) is the steepness of the sigmoid function,
    \item \(r_0\) characterizes the distance when fish starts reacting to predator.
\end{itemize}

The choice of parameters (Table \ref{parameters_lopez_predator}) is determined by the following criteria:

\begin{itemize}
    \item \(r_0\): This parameter is the distance at which the predator function attains half its maximum, and therefore sets the region over which the predator is felt. The predator is located in the middle of the stimuli compartment, at a distance of \(150\) mm from the closest point accessible to the agent (the partition) and up to \(540\) mm from the farthest corner of the school compartment. Values below \(150\) place the transition outside the accessible region, so that the predator function is uniformly small and the predator has no effect; values above \(450\) place it beyond almost the entire compartment, so that the function saturates at its maximum and the predator reduces to a constant-magnitude pull with no spatial structure.
    \item \(\beta\): This parameter controls the width of the transition of the sigmoid, which is of order \(1/\beta\): We therefore require \(1/\beta\) to be at most of the order of the compartment scale, giving \(\beta \geq 0.005\) mm (transition width \(200\) mm). Values of \(\beta \geq 0.1\) reduce the sigmoid to almost a step function, in which case the predator acts identically at all positions on one side of \(r_0\) and cannot distinguish between them.
    \item \(c_P\): We choose the range similar to \(c_f\): Values below \(10^{-2}\) render the predator negligible relative to alignment, reproducing the no-stimuli case; values above \(10^4\) make the predator dominant at all times.

\end{itemize}

All other parameter ranges are chosen the same way as in no-stimuli case.

\subsubsection{Range-based model}

For investigating a specific range-based model, we here chose the model proposed by Couzin \cite{couzin2002collective}.

\paragraph{Interaction zones.}
Time is divided into discrete \(\tau\) steps. The individual has ranges of
repulsion, alignment and attraction, which define the three candidate
directions (Equations \ref{couzin-repulsion}--\ref{couzin-attraction}):

\begin{align}
    \mathbf{d}_r(t+\tau) &= -\sum_{j \neq i}^{n_r} \frac{\mathbf{r}_{ij}}{|\mathbf{r}_{ij}|}
    \label{couzin-repulsion} \\[4pt]
    \mathbf{d}_o(t+\tau) &= \sum_{j=1}^{n_o} \mathbf{d}_j(t)
    \label{couzin-alignment} \\[4pt]
    \mathbf{d}_a(t+\tau) &= \sum_{j \neq i}^{n_a} \frac{\mathbf{r}_{ij}}{|\mathbf{r}_{ij}|}
    \label{couzin-attraction}
\end{align}

The neighbors in the zones of alignment or attraction are taken into account
only if they are in the field of perception of the fish (\(\alpha\) degrees).

\paragraph{Direction update.}
The repulsion has the highest priority, so if there is at least one neighbor in
that area, then \(\mathbf{d}_i(t+\tau)=\mathbf{d}_r(t+\tau)\). Otherwise, if
there are only fish either in alignment or attraction zone,
\(\mathbf{d}_i(t+\tau)=\mathbf{d}_o(t+\tau)\) and
\(\mathbf{d}_i(t+\tau)=\mathbf{d}_a(t+\tau)\) respectively. If there are
neighbors in both zones, then
\(\mathbf{d}_i(t+\tau)=\frac{1}{2}\left(\mathbf{d}_o(t+\tau)+\mathbf{d}_a(t+\tau)\right)\).
If there are no fish in any zones, or the resulting vector is zero, then
\(\mathbf{d}_i(t+\tau)=\mathbf{d}_i(t)\). Equivalently
(Equation \ref{couzin-update}),

\begin{equation}
    \mathbf{d}_i(t+\tau)=
    \begin{cases}
        \mathbf{d}_r(t+\tau), & n_r \geq 1, \\[2pt]
        \mathbf{d}_o(t+\tau), & n_r = 0, \; n_o \geq 1, \; n_a = 0, \\[2pt]
        \mathbf{d}_a(t+\tau), & n_r = 0, \; n_o = 0, \; n_a \geq 1, \\[2pt]
        \dfrac{1}{2}\left(\mathbf{d}_o(t+\tau)+\mathbf{d}_a(t+\tau)\right),
            & n_r = 0, \; n_o \geq 1, \; n_a \geq 1, \\[6pt]
        \mathbf{d}_i(t), & \text{otherwise.}
    \end{cases}
    \label{couzin-update}
\end{equation}

where,
\begin{itemize}
    \item \(\mathbf{r}_{ij}(t) = \mathbf{r}_j(t)-\mathbf{r}_i(t)\) is the
    relative position of the \(j^{\text{th}}\) fish with respect to the
    \(i^{\text{th}}\) fish,
    \item \(\mathbf{d}_j(t)\) is the directional unit vector of the
    \(j^{\text{th}}\) fish at the time \(t\),
    \item \(n_r\), \(n_o\) and \(n_a\) are the numbers of neighbors in the
    zones of repulsion, alignment and attraction, respectively,
    \item \(\alpha\) is the field of perception of the fish (degrees).
\end{itemize}

\paragraph{Noise and turning constraint.}
After that, stochastic effect is added: \(\mathbf{d}_i(t+\tau)\) is rotated by
a random angle taken from a Gaussian distribution with standard deviation
\(\sigma\). In addition, fish have a turning rate (\(\theta\) degrees per second), so it
cannot turn more than \(\theta \tau\) in one time step.

The choice of parameters (Table \ref{parameters_couzin}) is determined by following criteria:

\begin{itemize}

\item \(r_r\): The distance between two robots is \(32\sqrt{2} \approx 45.25\). Therefore, for the agent to be able to pass between them, we require
\(
r_r < 16\sqrt{2}
\).
Consequently, larger values of \(r_r\) are not considered.

\item \(\Delta r_o\): We choose a range extending to approximately twice the distance between two robots. Considering larger values is unnecessary, since at distances of approximately \(80\)--\(90\), when the agent is near the robotic school, all robots fall within the same range.

\item \(\Delta r_a\): Similarly, we choose a range extending to approximately twice the distance between the robots, as larger values are unnecessary for the same reason.

\item \(\alpha\): The range was proposed by the authors, who reasonably considered values below \(200\) to be inappropriate.

\item \(\theta\): Values below \(10\) are clearly inappropriate, as was also observed by the authors.

\item \(s\): The robots move at a constant speed of \(10\text{mm}/\text{s}\). We consider it reasonable to investigate values from half this speed up to twice this speed.

\item \(\sigma\): The range was proposed by the authors.
\end{itemize}

\subsubsection{New generalized cohesion model}

\begin{figure}[H]
    \centering
    \begin{subfigure}{0.5\textwidth}
        \includegraphics[width=1\linewidth]{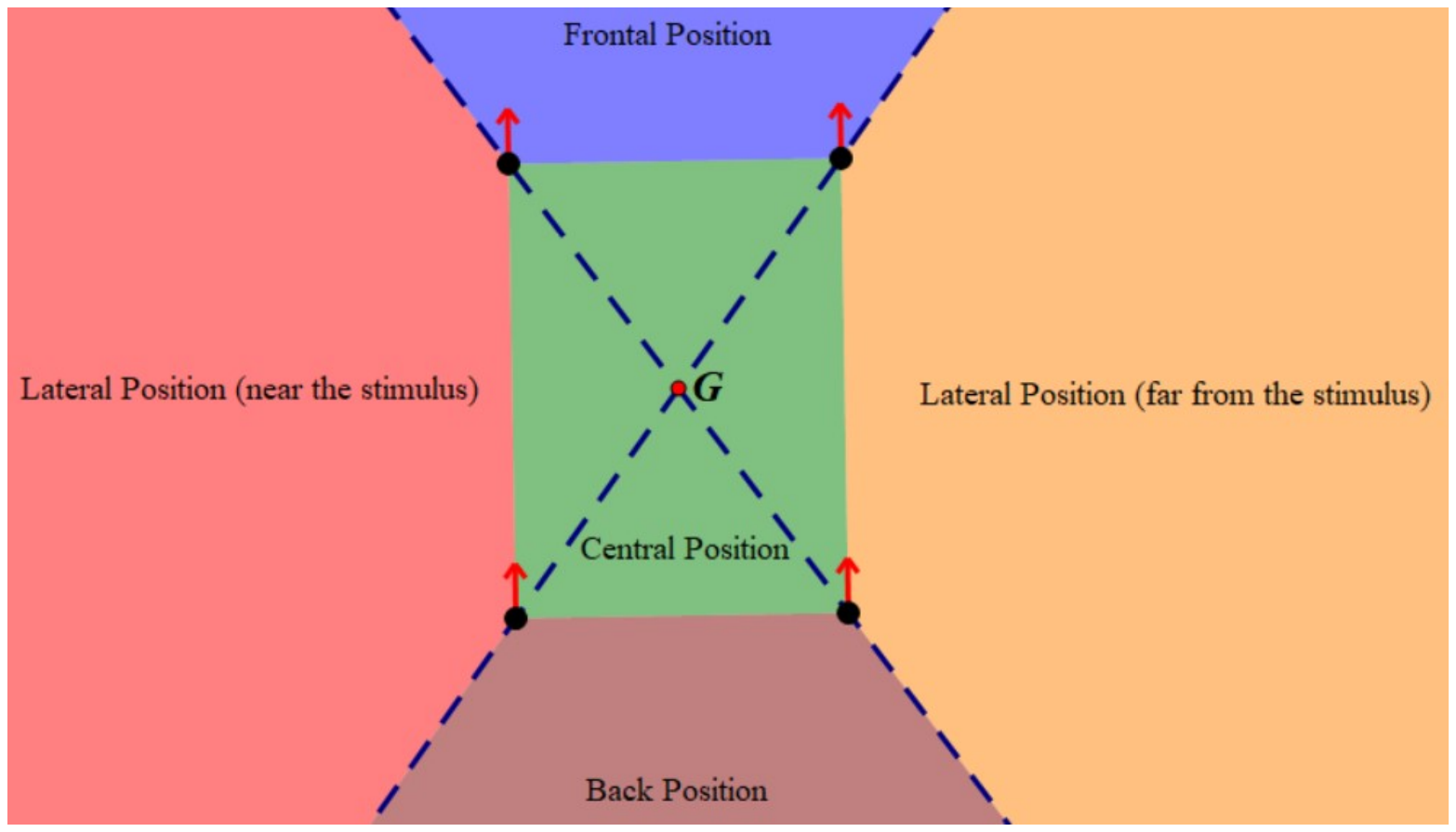}
    \caption{\footnotesize The definitions of the positions with respect to the four robotic fish (black dots in the picture) and their mass center \(G\).}
    \label{robotic_positions_definitions}
    \end{subfigure}
    
    \vspace{0.5cm}
    
    \begin{subfigure}{0.5\textwidth}
        \includegraphics[width=1\linewidth]{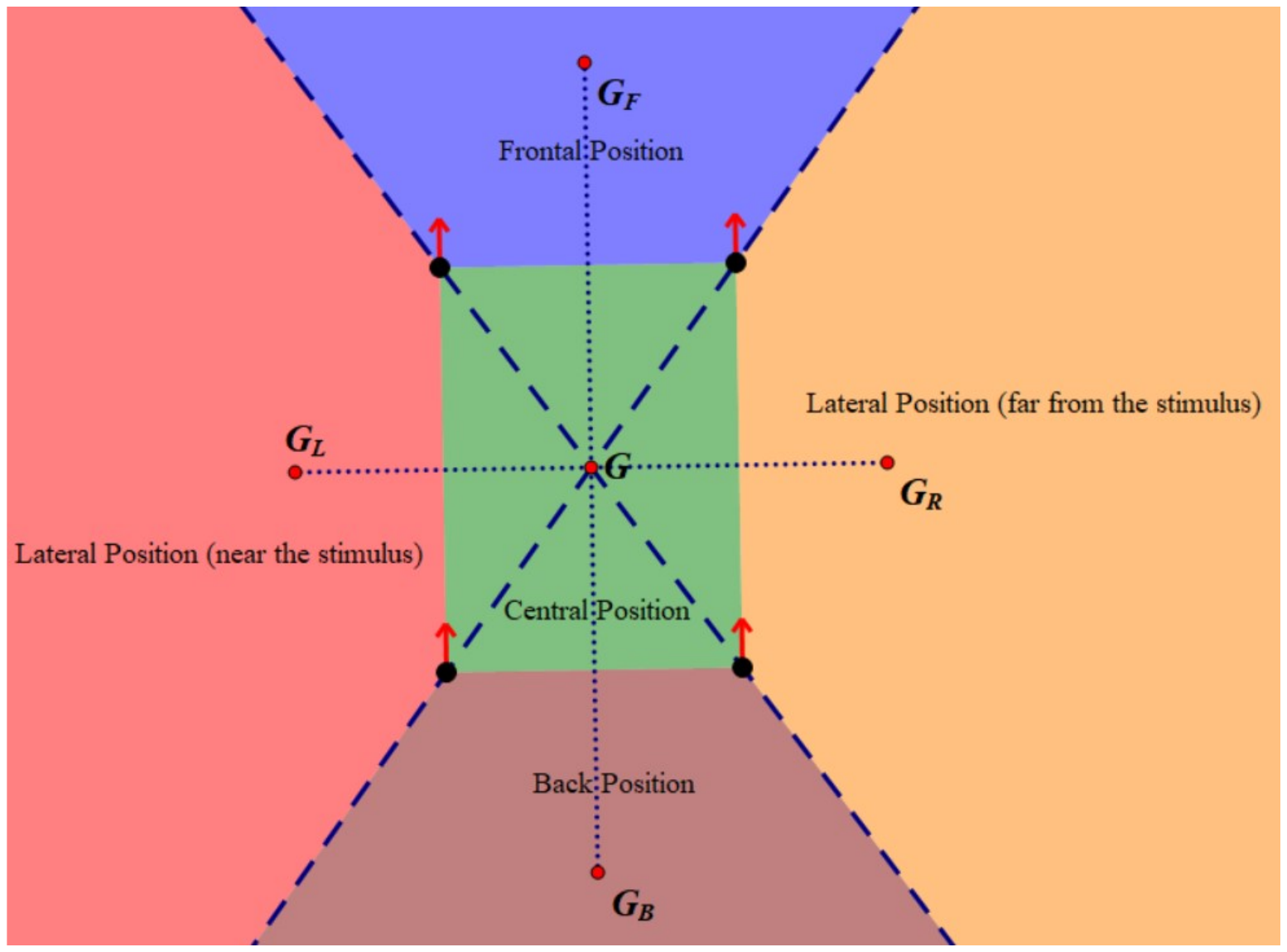}
    \caption{\footnotesize Definition of attraction points for the generalized cohesion model, with respect to the robotic school. The attraction points of the position force (\(G_L, G_F, G_R, G_B\)) are the reflections of \(G\) with respect to the sides of the rectangle}
    \label{Robotic fishes_position force}
    \end{subfigure}

    \vspace{0.5cm}

    \begin{subfigure}{0.5\textwidth}
        \includegraphics[width=1\linewidth]{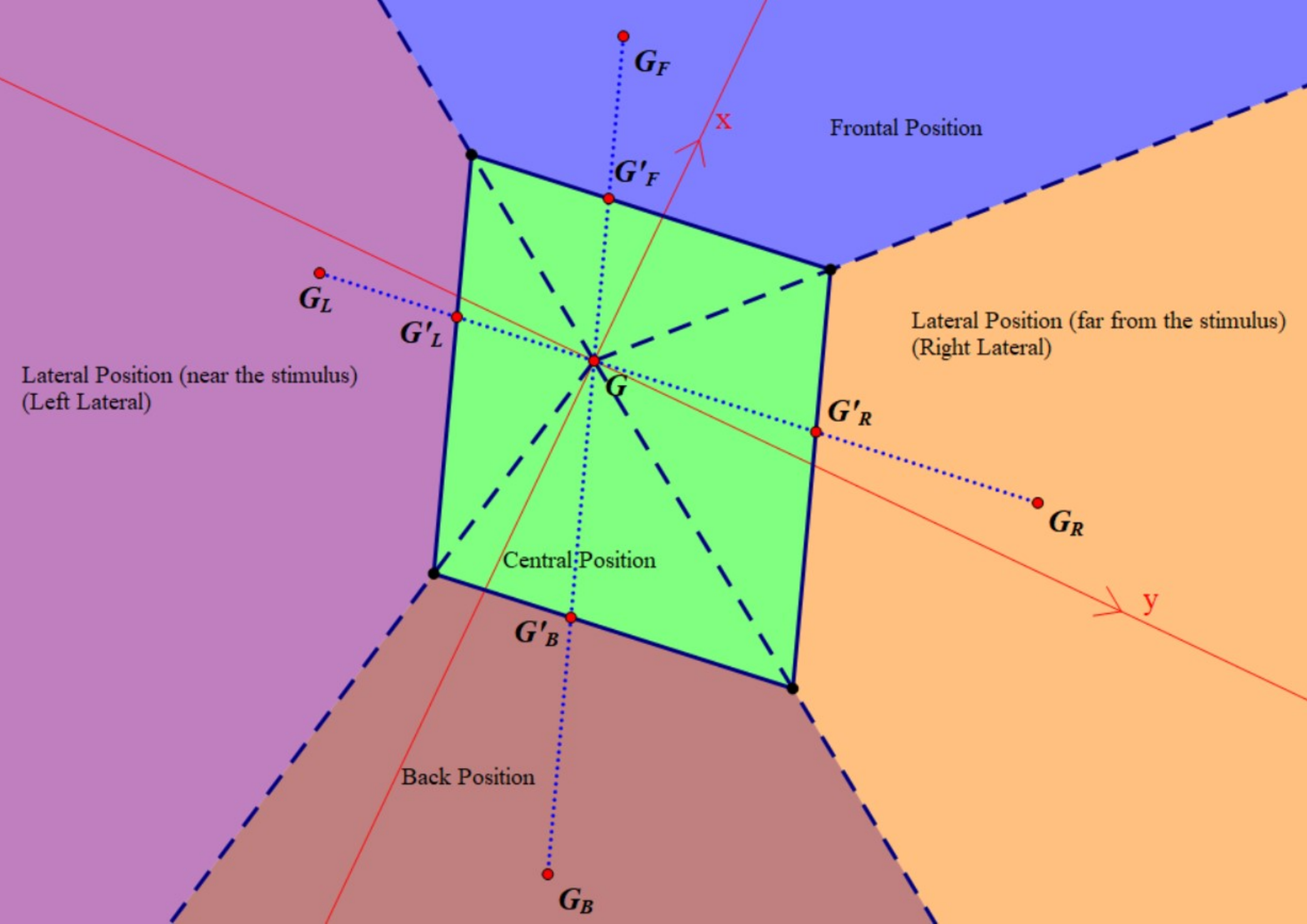}
    \caption{\footnotesize Generalized cohesion model: The definitions of positions in case of multiple fish}
    \label{Mathematical model_multiple fishes_positions definitions}
    \end{subfigure}
    
    \caption{Definitions of positions}
\end{figure}

In this proposed model, we introduce a new force called \textit{the position force} (Equation \ref{mathematical_model_equation}). As before, the fish moves at a constant speed \(s\). In addition to the previously described forces ~(Equation  \ref{Reynold_equation1}), each fish randomly selects a preferred position according to the experimentally derived probability distribution (Equation \ref{position_probabilities}), and the position force directs the fish toward that position. The force vector is directed toward the attraction point associated with the selected preferred position (Figure~\ref{Robotic fishes_position force}). Its magnitude equals the distance between the fish and the attraction point, so that \(p_i(t)=\mathbf{a}-\mathbf{r}_i(t)\). The weight \(c_p\) then scales this force relative to the others.

Once the fish has remained at the selected position for \(\Delta t\) seconds, it draws a new position and the process repeats. (Algorithm \ref{alg:mathematical_model}).

\begin{equation}
    \mathbf{d}_i(t+\tau) = c_w\frac{1}{n(t)} \sum_{j \in N_i(t)} w(|\mathbf{r}_{ij}(t)|) \mathbf{d}_j(t)
    + c_f\frac{1}{n(t)} \sum_{j \in N_i(t)} f(|\mathbf{r}_{ij}(t)|) \frac{\mathbf{r}_{ij}(t)}{|\mathbf{r}_{ij}(t)|} + c_p p_i(t) +\eta_i(t)
    \label{mathematical_model_equation}
\end{equation}

where,
\begin{itemize}
    \item \(w(r)\) is chosen to be constantly equal to \(1\).
    \item \(f(r)\) is the Lennard-Jones ~(\ref{Lennard_Jones_General}), as before.
    \item \(p_i(t)\) is the position force, directed from the fish toward the attraction point of its currently preferred position. The preferred position is drawn from the experimental distribution (Equation \ref{position_probabilities}) at initialization and again each time the fish has spent \(\Delta t = 10 \, \text{sec}\) at the position it selected.
    \item \(c_w, c_f\) and \(c_p\) are the weights of the alignment, attraction/repulsion and position forces respectively.
    \item \(\eta_i\) is a stochastic component (Gaussian with standard deviation \(\mu\)).
\end{itemize}

\begin{algorithm}[H]
\caption{Motion of the live fish under the generalized cohesion model}
\label{alg:mathematical_model}
\begin{algorithmic}[1]
\Require speed $s$, time step $\tau$, weights $c_w,c_f,c_p$, noise scale $\mu$,
         re-selection time $\Delta t$, distribution $p$ over the preferred positions $\mathcal{Z}$
\State $z\sim p(\cdot)$ \Comment{initial preferred position}
\For{$k=1,2,\dots,N$}
    \State $\mathbf{a}\gets$ attraction point of $z$ (Figure~\ref{Robotic fishes_position force})
    \State $\mathbf{d}\gets c_w\mathbf{W}+c_f\mathbf{F}+c_p(\mathbf{a}-\mathbf{r})$
           \Comment{alignment, attraction--repulsion~(\ref{Reynold_equation1}), position force}
    \State $\mathbf{d}\gets\mathbf{d}/\|\mathbf{d}\|+\boldsymbol{\eta},\qquad
           \eta_x,\eta_y\sim\mathcal{N}(0,\mu^{2})$
    \State $\mathbf{r}\gets\mathbf{r}+s\tau\,\mathbf{d}/\|\mathbf{d}\|$
           \Comment{constant speed $s$}
    \If{the fish has stayed at $z$ for $\Delta t$}
        \State $z\sim p(\cdot)$ \Comment{a new position is drawn, the process repeats}
    \EndIf
\EndFor
\end{algorithmic}
\end{algorithm}

This model is an extension of the generalized force-based model. Note that the \(c_p=0\) case is exactly the generalized force-based model. The aim of introducing the position force \(p_i(t)\) is to characterize the positional preference of the fish, and introduce exploration behavior besides exploitation.

We test the parameter ranges specified in Table \ref{parameters_mathematical}.

\paragraph{Generalizing to multiple fish}

To simulate a school consisting of multiple individuals using the proposed model, it is necessary to define the preferred positions of each individual within its neighborhood. We adopt a metric neighborhood, meaning that the neighborhood of an individual consists of all fish located within a distance of \(R\) units from it.
We define the positions as shown in Figure \ref{Mathematical model_multiple fishes_positions definitions}. \(G\) is the barycenter of the neighbors in the defined neighborhood. The red \(x\)-axis is the average direction of the neighbors, and the red \(y\)-axis is perpendicular to it. \(G'_F\) is the barycenter of the neighbors which are above \(y\)-axis, and \(G'_B\) is the barycenter of the ones below \(y\)-axis. In the same way, \(G'_L\) is the barycenter of the neighbors that are to the left of the \(x\)-axis, and \(G'_R\) is the barycenter of the ones that are to the right of the \(x\)-axis. The attraction points \(G_F, G_B, G_L, G_R\) of the positions are chosen to be the reflections of \(G\) with respect to \(G'_F, G'_B, G'_L, G'_R\) respectively. The attraction point of the \texttt{Central Position} is \(G\). It is worth noting that this definition is a generalization of the one introduced for the case of four robotic fish. When applied to that setting, the generalized definition reduces to the original one.

\subsection{Statistical Analysis}

\subsubsection{Simulation-Based Parameter Screening}

For a given model, we seek the parameter values that best reproduce the experimental observations. To this end, we adopt a procedure inspired by the Approximate Bayesian Computation (ABC) method
\cite{rubin1984bayesianly, beaumont2002approximate}; however, we use it as a screening tool to
identify parameter combinations compatible with the experimental data, rather than to estimate a posterior distribution.

The experimental dataset consists of \(20\) observations and \(5\) variables, where each column represents the time spent in the corresponding position across the \(20\) experiments (for a description of the experimental setup, refer to the \emph{Experimental Set-up} section \ref{SetUp}). The experimental data were scaled so that each row sums to $1200$. The same type of dataset is obtained by simulating a given schooling model. To quantify how well a particular model, with a specific set of parameters, reproduces the experimentally observed behavior, we define an error measure between the two datasets.

\paragraph{Error between simulation and experimental observations.}

\begin{itemize}
    \item Denote by \(X_\text{sim}\) the \(20\) vectors of total times spent in each position obtained by \(20\) independent simulations.
    \begin{equation*}
        X_\text{sim}=\begin{pmatrix}
            T_{1, 1} & T_{1, 2} & T_{1, 3} & T_{1, 4} & T_{1, 5} \\
            \vdots & & & & \vdots \\
            T_{20, 1} & T_{20, 2} & T_{20, 3} & T_{20, 4} & T_{20, 5}
        \end{pmatrix}
    \end{equation*}
    \item Denote by \(X_\text{obs}\) the experimental dataset with \(20\) observations.
\end{itemize}

So \(X_\text{sim}, X_\text{obs} \in \mathbb{R}^{20 \times 5}\). Define a distance (error) \(\rho\) on that space as follows
\begin{equation*}
    \begin{aligned}
        & \rho_j(X, X')=\frac{1}{400}\sum_{i=1}^{20} \sum_{i'=1}^{20} (X_{ij}-X'_{i'j})^2 & \text{(Error of one position)} \\
        & \rho(X, X')=\frac{1}{5}\sum_{j=1}^5 \rho_j(X, X') & \text{(Overall error)}
    \end{aligned}
\end{equation*}

Since every row of \(X\) is compared with every row of \(X'\), the error
compares the distributions of the times spent in each position, rather than
matched pairs of experiments. Consequently, \(\rho(X, X)\) is not necessarily
\(0\). We expect a well-tuned agent to produce an error close to
\(\rho(X_\text{obs}, X_\text{obs})\).

\paragraph{Parameter sampling and acceptance criterion.}

Denote by \(\boldsymbol{\vartheta}\) the vector of the model parameters. We
perform \(N=10^5\) experiments as follows:

\begin{enumerate}
    \item Generate \(\boldsymbol{\vartheta}_k\), \(k=1,\dots,N\), with every
    parameter sampled randomly from its specified distribution (i.e., uniform in the range, or log-uniform where specified).
    \item Run \(20\) independent simulations with every
    \(\boldsymbol{\vartheta}_k\), using an agent with the resulting combination of parameters, to obtain \(X_{\text{sim}, k}\).
    \item Choose a tolerance \(\delta\), and whenever
    \(\rho(X_{\text{sim}, k}, X_\text{obs})<\delta\), accept the sample \(\boldsymbol{\vartheta}_k\).
\end{enumerate}

In the no-stimuli case, under this
scaling, the error of the experimental data with itself is $7652.24$, and the
mean of the \emph{Frontal Position} column is $98.03$. For every parameter
combination we compute the error between the simulated and the experimental
data, and we expect a good fit to yield an error close to $7652.24$. We adopted a tolerance of $\delta =10000$, i.e.\ we consider only those parameter combinations whose error is below $10000$. Many of the runs satisfying this criterion, however, produced a \emph{Frontal Position} that is identically
zero. Since our main goal is to reproduce the exploration behaviour of
the fish, such cases are discarded: they fail to exhibit any exploration of the
frontal position. Such runs are therefore excluded from the tables below, although they are still counted among the accepted combinations.

In the predator stimuli case, under this
scaling, the error of the experimental data with itself is $4762.57$. For every parameter
combination we compute the error between the simulated and the experimental
data, and we expect a good fit to yield an error close to $4762.57$. We adopted a tolerance of $\delta=8000$, i.e.\ we consider only those parameter combinations whose error is below $8000$.

\subsubsection{Local Sensitivity Analysis}
\label{sec:local_sensitivity}

The screening procedure identifies parameter combinations whose error falls
below the tolerance $\delta$, but it does not show how robust these
combinations are. A biologically plausible behavioural rule should keep
producing similar behaviour under small changes of its parameters. To test
this for the models, we examine how the error changes in the
neighbourhood of a well-performing parameter combination. We reuse the
$N=10^5$ parameter combinations already simulated during the screening, so no
new simulations are needed.

\paragraph{Normalised parameter space.}
We consider only the parameters that were sampled during the screening.
Parameters held fixed ($\epsilon$, $c_w$, $\tau$) take the same value in
every combination and are therefore excluded. Because the sampled parameters
have different units and ranges, and some were drawn on a logarithmic scale,
we first map each one onto the unit interval. Let
$\boldsymbol{\vartheta}=(\vartheta_1,\dots,\vartheta_m)$ denote the vector of
sampled parameters. If $\vartheta_l$ was sampled uniformly on $[a_l,b_l]$, we
set
\begin{equation*}
    \tilde{\vartheta}_l = \frac{\vartheta_l - a_l}{b_l - a_l}.
\end{equation*}
If $\vartheta_l$ was sampled log-uniformly on $[10^{a_l},10^{b_l}]$, we set
\begin{equation*}
    \tilde{\vartheta}_l = \frac{\log_{10}\vartheta_l - a_l}{b_l - a_l}.
\end{equation*}
In both cases $\tilde{\vartheta}_l\in[0,1]$, and the bounds $a_l,b_l$ are the
nominal ranges of the screening. For the generalized force-based models
(constant and positive Lennard--Jones alignment), the sampled parameters are
$\sigma$, $\alpha$, $s$ and $\mu$, which are uniform, and $c_f$, which is
log-uniform. This gives $m=5$ (Table~\ref{parameters_lopez}). For the range-based
model, the sampled parameters $r_r$, $\Delta r_o$, $\Delta r_a$, $\alpha$,
$\theta$, $s$ and $\sigma$ are all uniform, which gives $m=7$
(Table~\ref{parameters_couzin}). For the generalized cohesion model, the sampled parameters are $\sigma$,
$\alpha$, $s$ and $\mu$, which are uniform, and $c_f$ and $c_p$, which are log-uniform. This gives $m=6$ (Table~\ref{parameters_mathematical}). In the predator context, the generalized force-based model has three additional sampled parameters: $r_0$ and $\beta$, which are uniform, and $c_P$, which is log-uniform, giving $m=8$ (Table~\ref{parameters_lopez_predator}).

\paragraph{Distance between parameter combinations.}
The distance between two parameter combinations $\boldsymbol{\vartheta}_k$
and $\boldsymbol{\vartheta}_{k'}$ is the Euclidean distance between their
normalised vectors:
\begin{equation*}
    D(\boldsymbol{\vartheta}_k,\boldsymbol{\vartheta}_{k'})
    = \left\| \tilde{\boldsymbol{\vartheta}}_k - \tilde{\boldsymbol{\vartheta}}_{k'} \right\|_2
    = \sqrt{\sum_{l=1}^{m}\left(\tilde{\vartheta}_{k,l}-\tilde{\vartheta}_{k',l}\right)^2 }.
\end{equation*}

\paragraph{Neighbourhood of a well-performing combination.}
For each model, we select as reference a combination
$\boldsymbol{\vartheta}_{k^\ast}$ that was accepted by the screening and
reproduces the experimental distributions well. We compute
$D(\boldsymbol{\vartheta}_{k^\ast},\boldsymbol{\vartheta}_k)$ for all
$k\neq k^\ast$ among the $10^5$ screened combinations and retain the five
with the smallest distance, i.e.\ the $5$ nearest neighbours of the
reference in the normalised parameter space. For the reference and its
neighbours, we report the parameter values, the distance $D$, the error $\rho$ and the Frontal Position mean (no-stimuli case), together with the performance of the reference combination (Table~\ref{tab:knn_all}).

\subsubsection{Statistical Measures}
\label{measures}

We will walk through several statistical measures for investigating features of a school.

The \textit{polarization} of the school is defined as follows:

\begin{equation*}
    P(t)=\frac{1}{N} \left| \sum_{i=1}^N d_i(t) \right|
\end{equation*}

where,
\begin{itemize}
    \item \(N\) is the number of individuals in the school,
    \item \(d_i(t)\) is the directional unit vector of the \(i^\text{th}\) individual at time \(t\).
\end{itemize}

The polarization always satisfies \(0 \leq P(t) \leq 1\). Assuming the individuals form a single school, the polarization is close to \(0\) when the school exhibits swarming behavior and approaches \(1\) when the school displays parallel behavior, with the individuals moving in a common direction.

We further define \textit{average direction} of the school as the direction of the vector \(\displaystyle\frac{1}{N}\sum_{i=1}^N d_i(t)\).

Subsequently, we define \textit{mean} and \textit{max polarization} of a school over \(T\) time steps as
\begin{equation*}
    \begin{aligned}
        & P_\text{mean} = \frac{1}{T}\sum_{t=1}^T P(t) \\
        & P_\text{max}=\max_{1 \leq t \leq T} P(t)
    \end{aligned}
\end{equation*}

We consider the rectangle of minimum area such that its sides are parallel and perpendicular to the average direction of the school, with all individuals contained within the rectangle. We define the \textit{length} of the school as the length of the side parallel to the average direction, and the \textit{width} as the length of the side perpendicular to it \cite{kunz2003artificial}.

\section{Results}
\label{results}

\subsection{Artificial School Experiment}

Here we report the results for the experiments with four robotic fish and one simulated agent.

\subsubsection{Generalized force-based model: Constant Alignment}

\begin{figure}[H]
    \centering
    \begin{subfigure}[t]{0.48\textwidth}
        \includegraphics[width=\textwidth]{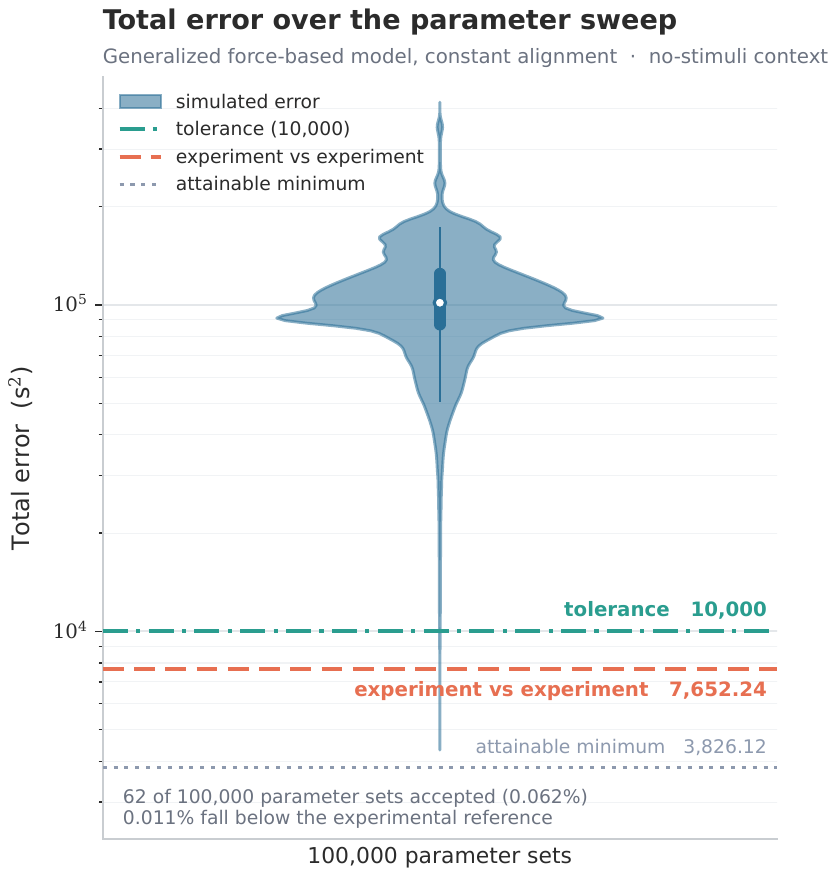}
        \caption{\footnotesize Generalized force-based model, constant alignment}
        \label{Lopez model_ABC_error_total}
    \end{subfigure}
    \hfill
    \begin{subfigure}[t]{0.48\textwidth}
        \includegraphics[width=\textwidth]{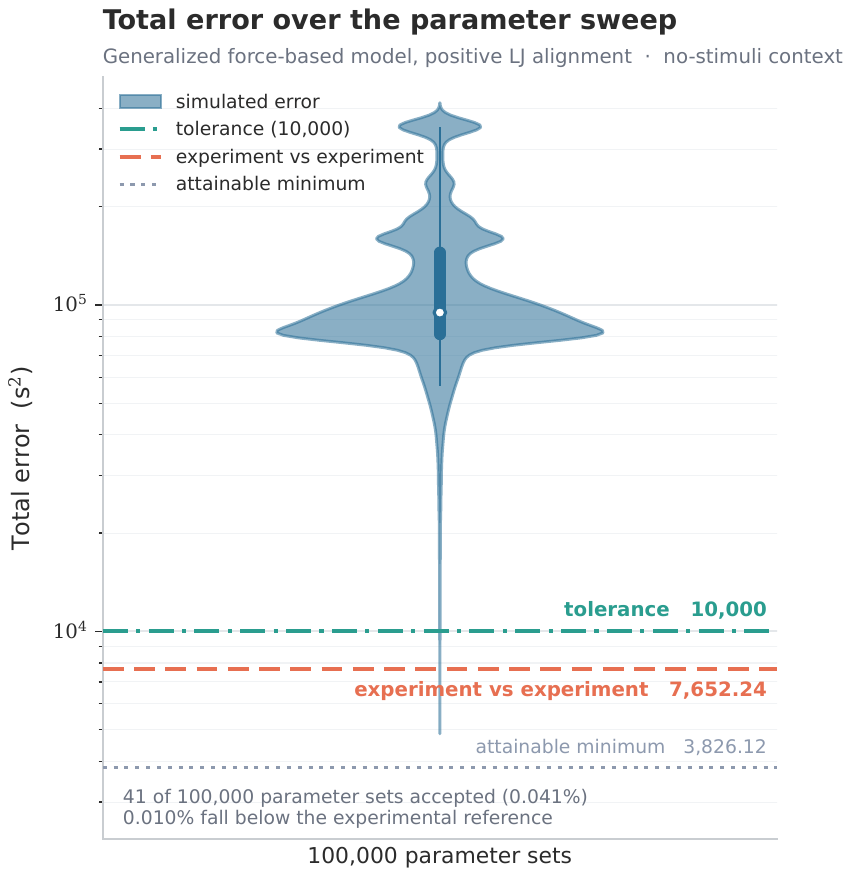}
        \caption{\footnotesize Generalized force-based model, positive LJ alignment}
        \label{Lopez model_other_ABC_error_total}
    \end{subfigure}
    
    \vspace{0.5cm}
    
    \begin{subfigure}[t]{0.48\textwidth}
        \includegraphics[width=\textwidth]{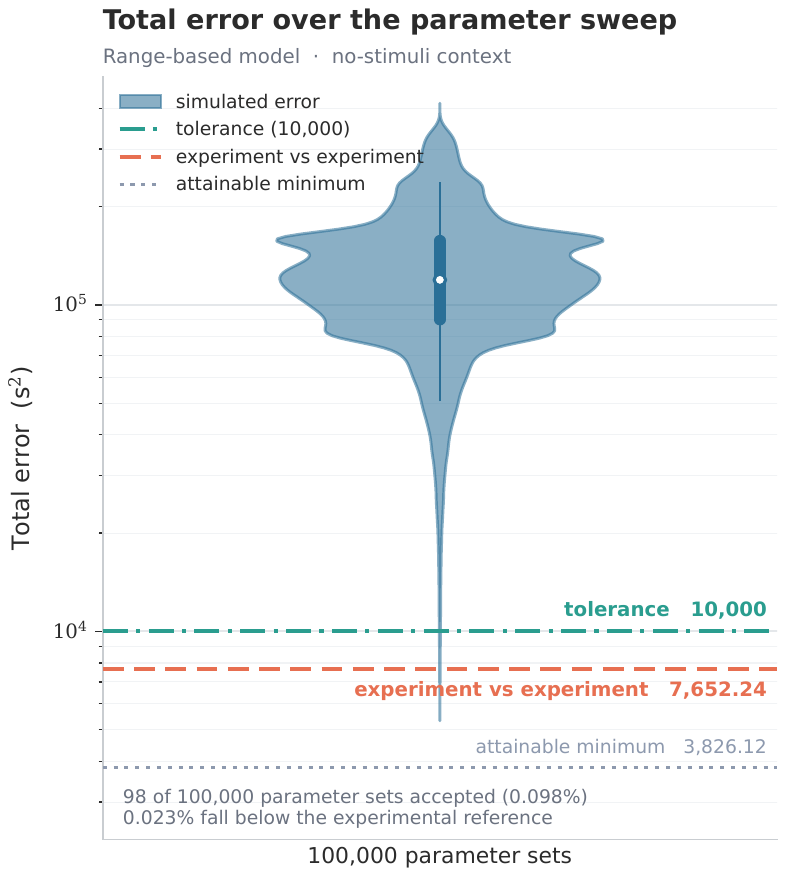}
        \caption{\footnotesize The range-based model}
        \label{Couzin model_ABC_error_total}
    \end{subfigure}
    \hfill
    \begin{subfigure}[t]{0.48\textwidth}
        \includegraphics[width=\textwidth]{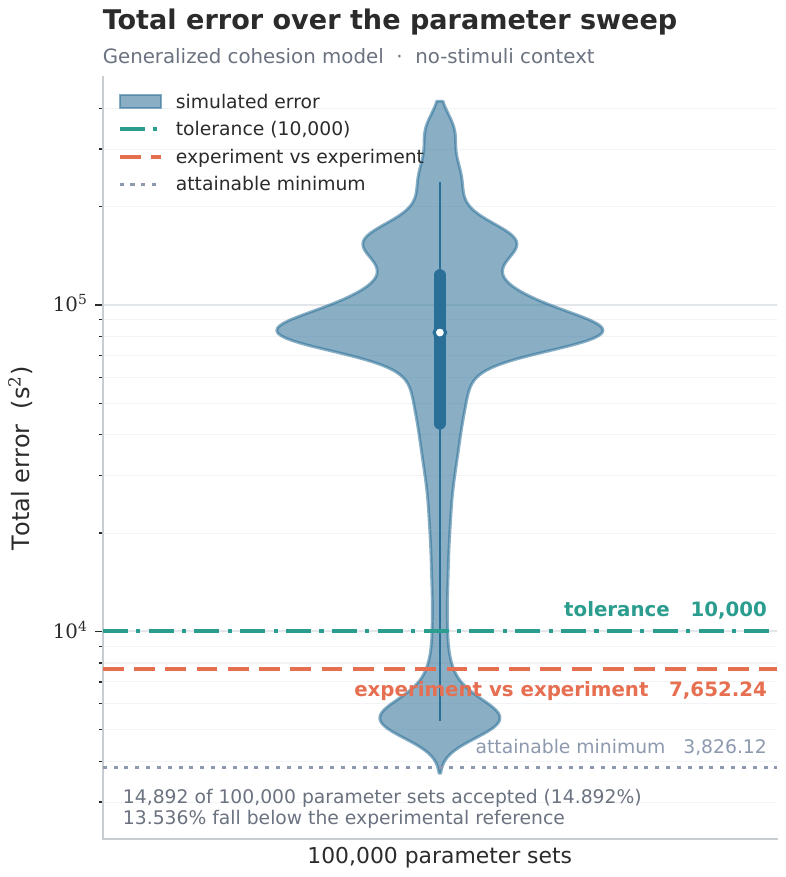}
        \caption{\footnotesize Generalized cohesion model}
        \label{Mathematical model_ABC_error_total}
    \end{subfigure}
    \caption{The distribution of the error across the \(100000\) parameter combinations}
    \label{ABC_error_total}
\end{figure}

The results of the full experiment are shown in Figure \ref{Lopez model_ABC_error_total}. We see that only \(62\) parameter combinations out of \(10^5\) fell below the tolerance.

We further investigate the cases with error below the tolerance rate and non-zero ``Frontal Position" mean (Table \ref{tab:lopez-const}). We obtained nine cases here. Cases~5, 6 and~7 had a very low
$\sigma$. Recalling that the distance at which the repulsion force vanishes is
$\sigma\,2^{1/\alpha}$, this distance is very small; inspecting the animations
produced with these parameters, we observed a strange, non-fish-like behaviour
in which the agents spin rapidly around the robots. These three cases can
therefore be argued to be insufficient, as they do not represent plausible fish behaviour.

The other cases do not suffer from this problem. The only objection that is possible to raise against them is that their Frontal Position mean is far below the
experimental value of $98.03$, whereas that of case~8 is far above it.

Table~\ref{tab:knn_lopez_const} shows case~1 together with its five nearest
neighbours in the normalised parameter space
(Section~\ref{sec:local_sensitivity}). Four of the five neighbours have an
error far above the tolerance. The remaining one (run 12117) falls below the
tolerance, but its Frontal Position mean is zero, so it does not meet our
criterion.

\begin{table}[H]
\centering
\caption{Local sensitivity analysis in the no-stimuli context. For each model,
the reference combination (bold) and its five nearest neighbours in the
normalised parameter space are shown, ordered by distance $D$.}
\label{tab:knn_all}

\begin{subtable}{\textwidth}
\centering
\caption{Generalized force-based model, constant alignment, around case~1 of
Table~\ref{tab:lopez-const}. $\epsilon=1$ and $c_w=1$ in all rows.}
\label{tab:knn_lopez_const}
\begin{adjustbox}{max width=\textwidth}
\footnotesize
\begin{tabular}{lrrrrrrrr}
\toprule
Run ID & $\sigma$ & $\alpha$ & $c_f$ & $s$ & $\mu$ & $D$ & Error & FP mean \\
\midrule
\textbf{19179} & \textbf{15.80} & \textbf{2.37} & \textbf{58.49} & \textbf{9.63} & \textbf{0.26} & \textbf{0} & \textbf{7612.32} & \textbf{6.73} \\
23211 & 16.19 & 2.39 &  86.79 &  8.60 & 0.26 & 0.077 & 68313.16 &   0.00 \\
73119 & 14.72 & 2.53 &  52.55 & 10.23 & 0.29 & 0.081 & 58147.69 &   0.00 \\
12117 & 16.24 & 2.37 & 171.17 &  9.71 & 0.29 & 0.085 &  9062.85 &   0.00 \\
64779 & 14.82 & 2.32 &  74.97 &  8.79 & 0.31 & 0.089 & 22770.24 &  19.80 \\
77132 & 16.97 & 2.43 &  19.66 &  9.43 & 0.24 & 0.104 & 53580.28 & 247.07 \\
\midrule
\textbf{Experiment} & --- & --- & --- & --- & --- & --- & \textbf{7652.24} & \textbf{98.03} \\
\bottomrule
\end{tabular}
\end{adjustbox}
\end{subtable}

\vspace{0.4cm}

\begin{subtable}{\textwidth}
\centering
\caption{Generalized force-based model, positive Lennard--Jones alignment,
around case~3 of Table~\ref{tab:lopez-lj}. $\epsilon=1$ and $c_w=1$ in all rows.}
\label{tab:knn_lopez_lj}
\begin{adjustbox}{max width=\textwidth}
\footnotesize
\begin{tabular}{lrrrrrrrr}
\toprule
Run ID & $\sigma$ & $\alpha$ & $c_f$ & $s$ & $\mu$ & $D$ & Error & FP mean \\
\midrule
\textbf{82633} & \textbf{7.93} & \textbf{5.06} & \textbf{0.85} & \textbf{11.07} & \textbf{0.07} & \textbf{0} & \textbf{5726.12} & \textbf{111.71} \\
34784 & 7.54 & 5.07 & 0.84 & 11.26 & 0.02 & 0.056 &  13315.72 &    0.00 \\
40753 & 8.50 & 5.21 & 1.36 & 11.42 & 0.09 & 0.063 & 352729.22 & 1200.00 \\
54660 & 8.04 & 5.12 & 0.52 & 10.07 & 0.09 & 0.080 &  75073.41 &  416.65 \\
31571 & 7.42 & 5.24 & 0.76 & 11.71 & 0.01 & 0.090 &  89030.06 &    0.00 \\
47815 & 6.73 & 4.93 & 0.59 & 10.55 & 0.03 & 0.091 &  70084.98 &  401.56 \\
\midrule
\textbf{Experiment} & --- & --- & --- & --- & --- & --- & \textbf{7652.24} & \textbf{98.03} \\
\bottomrule
\end{tabular}
\end{adjustbox}
\end{subtable}

\vspace{0.4cm}

\begin{subtable}{\textwidth}
\centering
\caption{Range-based model, around case~1 of Table~\ref{tab:couzin}. The noise
amplitude $\sigma$ is reported in units of $10^{-2}$.}
\label{tab:knn_couzin}
\begin{adjustbox}{max width=\textwidth}
\footnotesize
\begin{tabular}{lrrrrrrrrrr}
\toprule
Run ID & $r_r$ & $\Delta r_o$ & $\Delta r_a$ & $\alpha$ & $\theta$ & $s$ &
$\sigma\,(\times 10^{-2})$ & $D$ & Error & FP mean \\
\midrule
\textbf{2606} & \textbf{22.83} & \textbf{8.84} & \textbf{29.87} & \textbf{324.83} & \textbf{86.31} & \textbf{11.92} & \textbf{6.89} & \textbf{0} & \textbf{7014.83} & \textbf{87.48} \\
2939  & 23.85 & 12.36 & 30.74 & 336.47 & 109.72 & 11.87 & 7.10 & 0.167 &  87255.42 &  19.38 \\
40948 & 20.05 & 18.29 & 33.73 & 321.69 &  84.21 & 11.23 & 6.74 & 0.170 & 104566.56 & 247.42 \\
14955 & 22.18 &  8.69 & 44.76 & 325.07 &  83.19 & 11.91 & 6.10 & 0.173 &  28003.76 & 258.39 \\
46025 & 23.06 & 22.63 & 34.31 & 334.49 &  98.41 & 12.71 & 6.48 & 0.195 &  58670.01 & 331.61 \\
67504 & 21.28 & 11.83 & 16.72 & 342.35 &  90.51 & 12.33 & 7.41 & 0.202 & 151094.35 & 781.87 \\
\midrule
\textbf{Experiment} & --- & --- & --- & --- & --- & --- & --- & --- & \textbf{7652.24} & \textbf{98.03} \\
\bottomrule
\end{tabular}
\end{adjustbox}
\end{subtable}

\vspace{0.4cm}

\begin{subtable}{\textwidth}
\centering
\caption{Generalized cohesion model, around run 4898. $\epsilon=1$ and $c_w=1$
in all rows.}
\label{tab:knn_mathematical}
\begin{adjustbox}{max width=\textwidth}
\footnotesize
\begin{tabular}{lrrrrrrrrr}
\toprule
Run ID & $\sigma$ & $\alpha$ & $c_f$ & $c_p$ & $s$ & $\mu$ & $D$ & Error & FP mean \\
\midrule
\textbf{4898} & \textbf{15.73} & \textbf{2.58} & \textbf{22.36} & \textbf{0.77} & \textbf{18.15} & \textbf{0.10} & \textbf{0} & \textbf{5112.35} & \textbf{84.19} \\
64916 & 17.82 & 2.51 & 31.14 & 0.65 & 18.48 & 0.01 & 0.147 & 5462.91 & 86.44 \\
9146  & 17.84 & 2.42 & 66.27 & 0.80 & 17.39 & 0.11 & 0.149 & 4943.81 & 85.10 \\
16375 & 14.44 & 2.47 & 89.23 & 0.85 & 18.10 & 0.19 & 0.155 & 5563.85 & 80.57 \\
2070  & 15.33 & 2.73 & 28.58 & 0.49 & 19.14 & 0.22 & 0.156 & 5073.40 & 95.21 \\
7963  & 14.76 & 3.12 & 55.20 & 0.70 & 19.08 & 0.15 & 0.157 & 5344.74 & 85.24 \\
\midrule
\textbf{Experiment} & --- & --- & --- & --- & --- & --- & --- & \textbf{7652.24} & \textbf{98.03} \\
\bottomrule
\end{tabular}
\end{adjustbox}
\end{subtable}
\end{table}

\begin{figure}[H]
\centering
\begin{subfigure}[t]{0.48\textwidth}
    \centering
    \includegraphics[width=\textwidth]{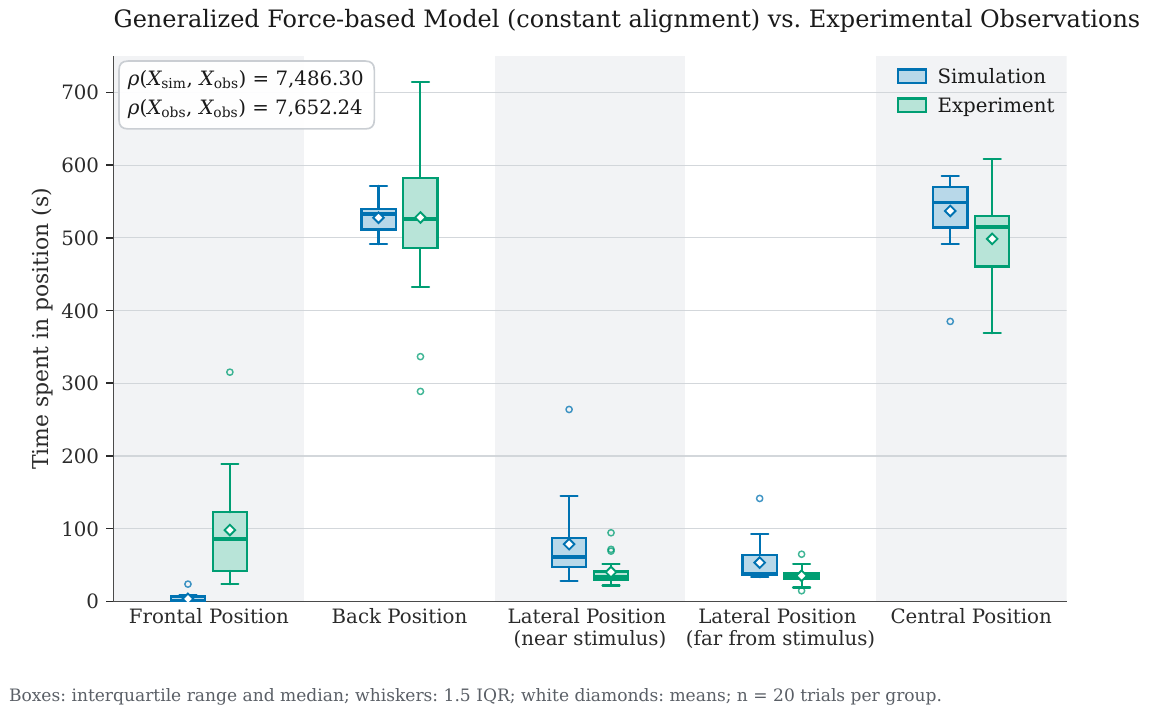}
    \caption{\footnotesize Generalized force-based model, constant alignment
    (run 19179).}
    \label{fig:lopez-const}
\end{subfigure}
\hfill
\begin{subfigure}[t]{0.48\textwidth}
    \centering
    \includegraphics[width=\textwidth]{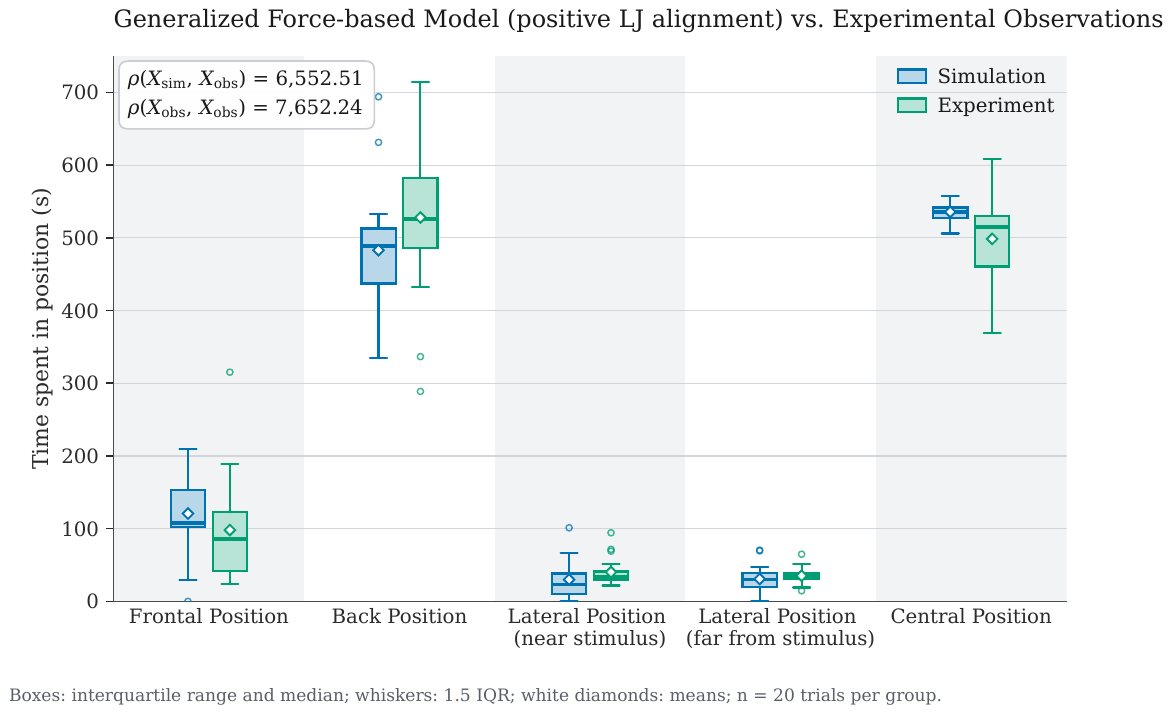}
    \caption{\footnotesize Generalized force-based model, positive
    Lennard--Jones alignment (run 82633).}
    \label{fig:lopez-lj}
\end{subfigure}

\vspace{0.5cm}

\begin{subfigure}[t]{0.48\textwidth}
    \centering
    \includegraphics[width=\textwidth]{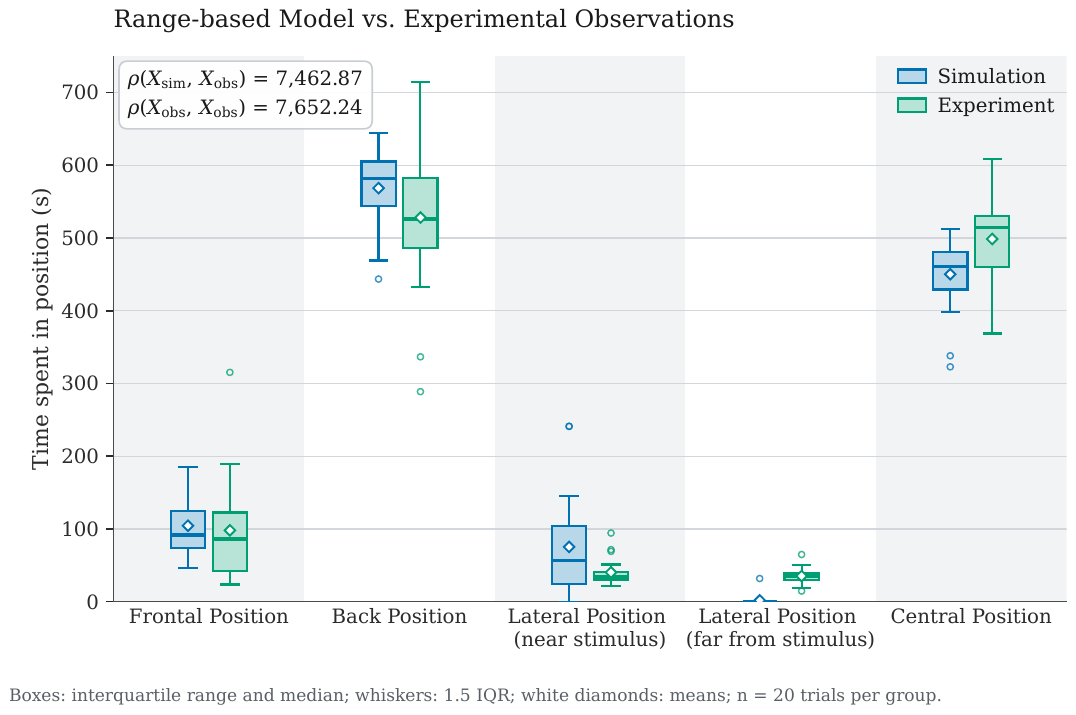}
    \caption{\footnotesize Range-based model (run 2606).}
    \label{fig:couzin1}
\end{subfigure}
\hfill
\begin{subfigure}[t]{0.48\textwidth}
    \centering
    \includegraphics[width=\textwidth]{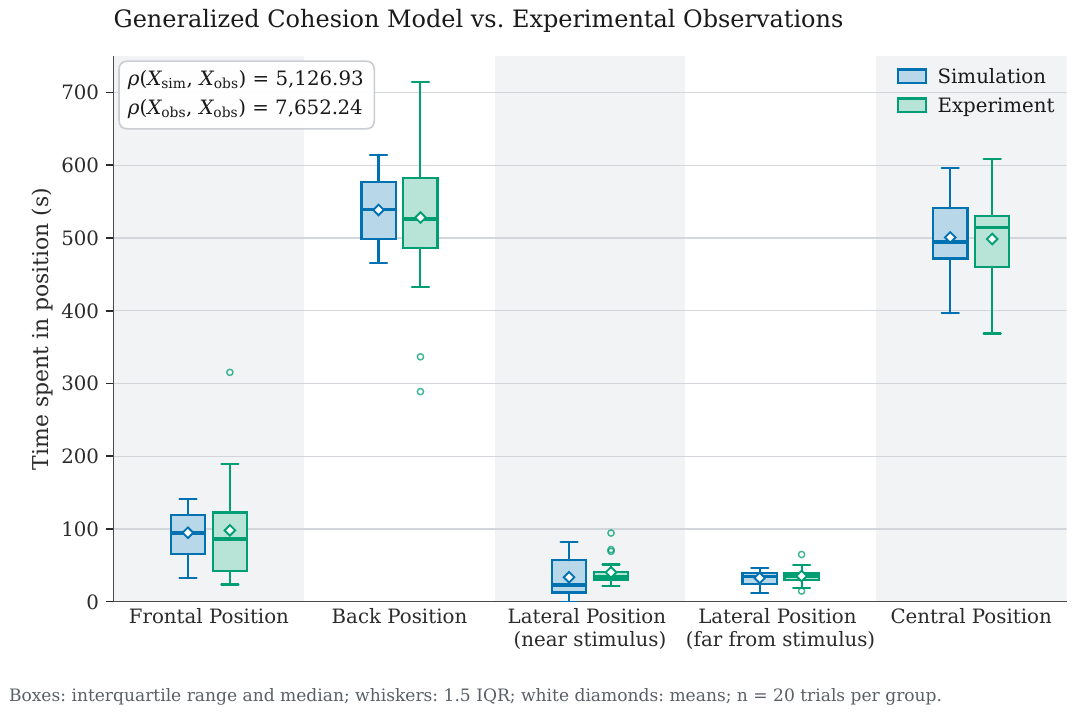}
    \caption{\footnotesize Generalized cohesion model (run 4898).}
    \label{fig:mathematical-perf}
\end{subfigure}
\caption{Performance of the reference combinations of
Table~\ref{tab:knn_all}, each simulated with a different seed, in the
no-stimuli context.}
\label{fig:knn_performance}
\end{figure}

\paragraph{Predator context:}

The results of the full experiment are shown in the Section \ref{SupplementaryMaterials}, as supplementary material. In Figure \ref{Lopez model_predator_ABC_error_total}, we see that only \(182\) parameter combinations out of \(10^5\) fell below the tolerance.

Performance of one of those is illustrated in Figure \ref{Lopez model_predator_performance_error}. As we can see, the agent is able to closely replicate the empirical distributions. Figure~\ref{fig:knn_lopez_predator} shows this combination (run 11442) together with its five nearest neighbours in the normalised parameter space (Section~\ref{sec:local_sensitivity}). As in the no-stimuli context, all five neighbours have an error far above the tolerance, the smallest being more than seven times that of the reference.

\subsubsection{Generalized force-based model: Positive Lennard-Jones Alignment}

The results of the full experiment are shown in Figure \ref{Lopez model_other_ABC_error_total}. We see that only \(41\) parameter combinations out of \(10^5\) fell below the tolerance.

We further investigate the cases with error below the tolerance rate and non-zero ``Frontal Position" mean (Table \ref{tab:lopez-lj}).

Only four runs passed the tolerance and non-zero ``Frontal Position" guard. Cases~1, 2 and~4 can be regarded as
insufficient because of their low Frontal Position mean; case~1 additionally
has a very low $\sigma$.

Case~3, on the other hand, performs rather well. Table~\ref{tab:knn_lopez_lj} shows case~3 together with its five nearest
neighbors. All five neighbors have an error above the tolerance, and in
four of them the error is more than ten times that of the reference.

\subsubsection{The range-based model}

The results of the full experiment are shown in Figure \ref{Couzin model_ABC_error_total}. We see that only \(98\) parameter combinations out of \(10^5\) fell below the tolerance.

We further investigate the cases with error below the tolerance rate and non-zero ``Frontal Position" mean (Table \ref{tab:couzin}).

While most of the rows in Table~\ref{tab:couzin} exhibit a rather low Frontal
Position mean, several of them meet our criteria. We will investigate the case~1, as it has the smallest error in the table. 
Table~\ref{tab:knn_couzin} shows case~1 together with its five nearest
neighbours. All five neighbours have an error well above the tolerance, and
their Frontal Position means range from $19.38$ to $781.87$.

\subsubsection{Generalized cohesion model}

The results of the full experiment are shown in Figure \ref{Mathematical model_ABC_error_total}. We see that \(14892\) parameter combinations out of \(10^5\) fell below the tolerance, which is significantly higher than the other models. We also observed that out of the \(14892\) accepted combinations, \(14588\) also have a non-zero Frontal Position mean.

We picked an accepted run to investigate (Figure \ref{fig:mathematical-perf}). Table~\ref{tab:knn_mathematical} shows run 4898 together with its five
nearest neighbours in the normalised parameter space. 
In contrast to the classical models,
all five neighbours have an error below the tolerance, and their Frontal
Position means remain close to the experimental value.

\subsection{Generalized cohesion model for multiple fish}

We wish to investigate the properties of the generalized cohesion model, in case of multiple live fish.

A total of \(N\) fish are initialized with positions sampled uniformly at random within a circle of radius \(150\). Their initial direction vectors are also chosen uniformly at random. In addition, each fish is assigned an initial time perception uniformly at random from \(\{1, \dots, \Delta t\}\), ensuring that individuals update their preferred positions at different time steps.

\begin{figure}[H]
    \centering
    \begin{subfigure}{\textwidth}
        \centering
        \includegraphics[width=\textwidth,height=0.36\textheight,keepaspectratio]{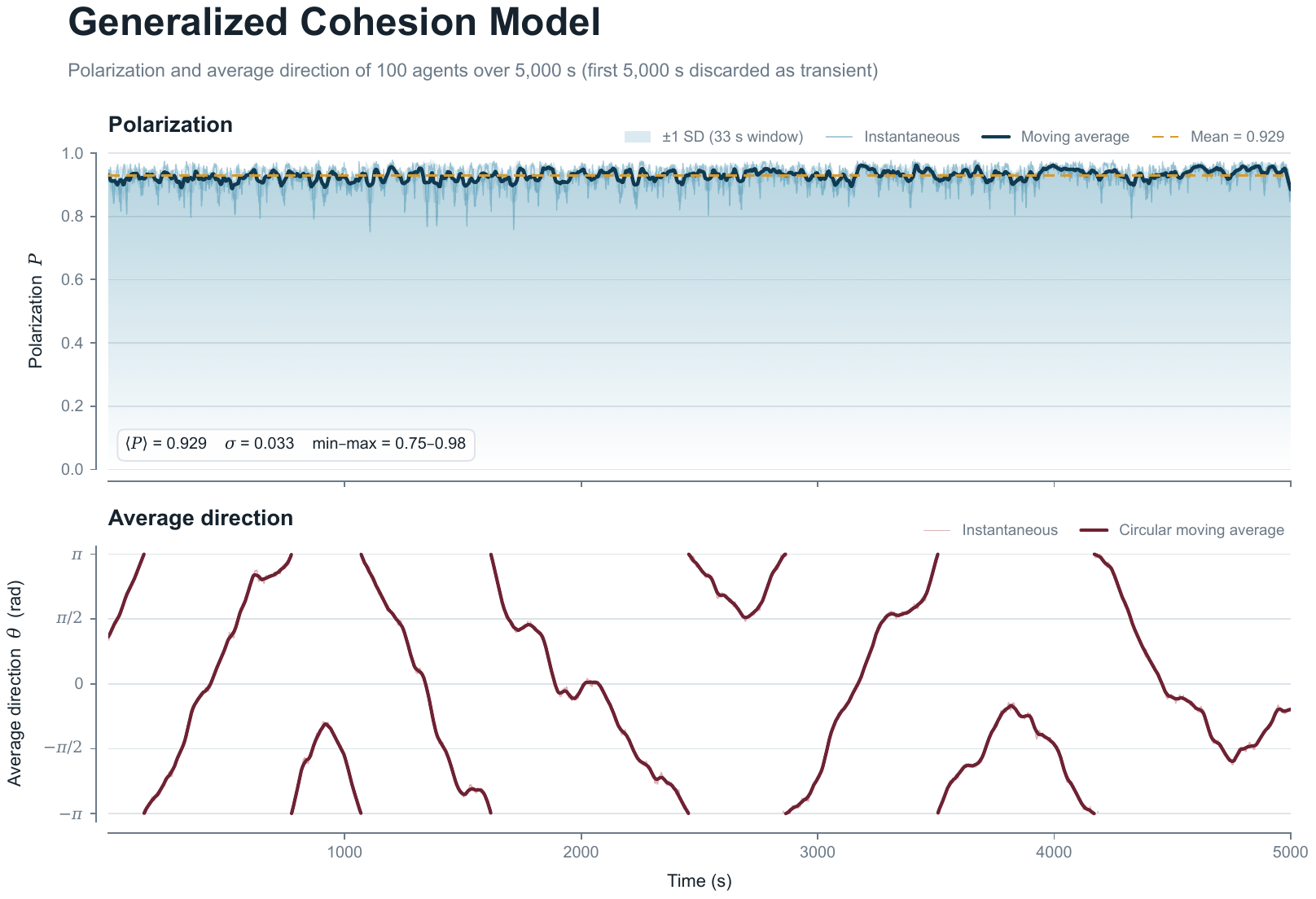}
        \caption{\footnotesize The change of polarization and average direction over time, with \(c_w=0.8\) and \(c_f=0.4\). The results are taken after the simulation of \(5000\) seconds, in order for the system to stabilize.}
        \label{Polarization_mathematical_1}
    \end{subfigure}

    \vspace{0.4cm}

    \begin{subfigure}{\textwidth}
        \centering
        \includegraphics[width=\textwidth,height=0.36\textheight,keepaspectratio]{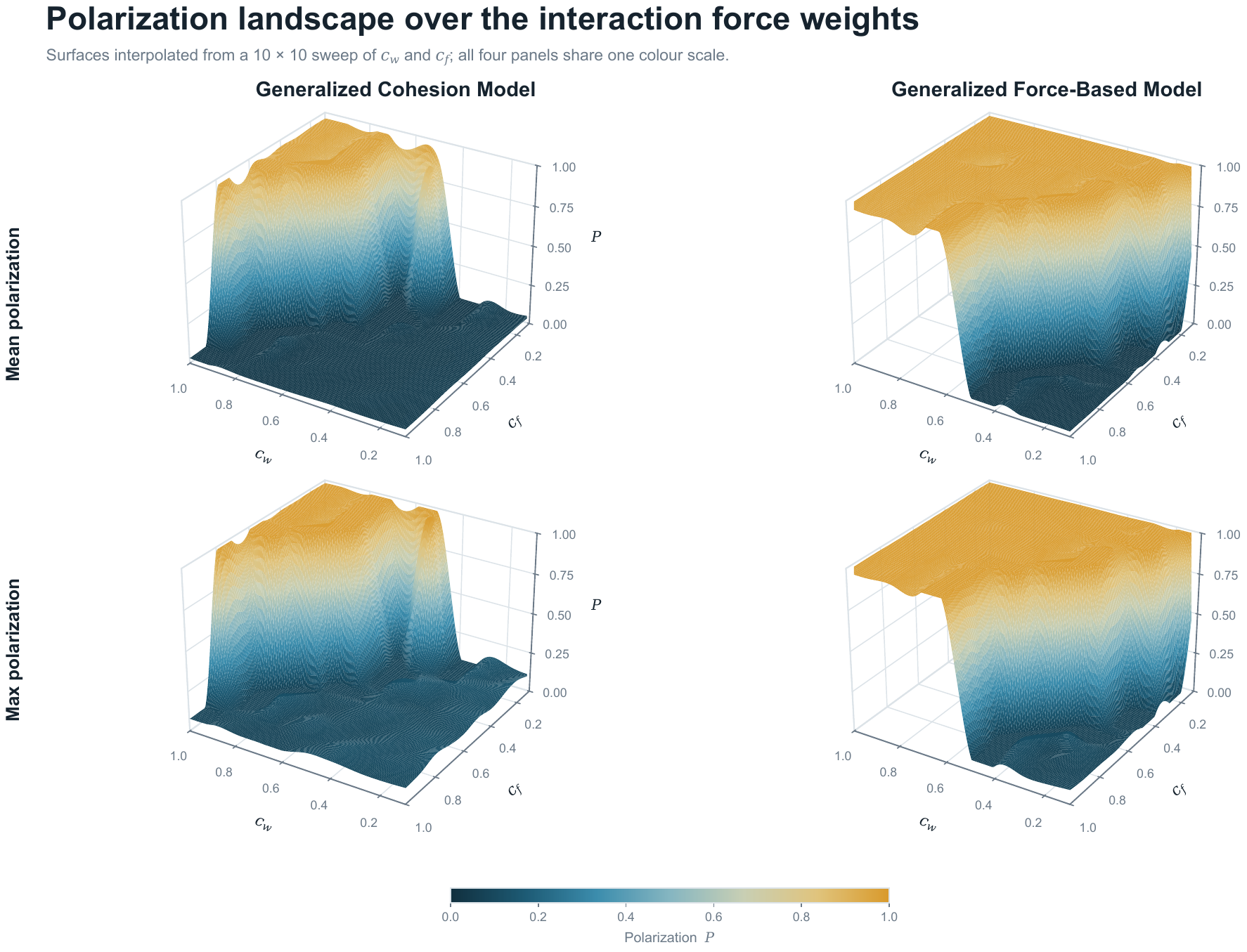}
        \caption{\footnotesize Generalized cohesion model compared to generalized force-based model: mean and max polarization vs weights. For each combination of \(c_w\) and \(c_f\), the school was first simulated for \(5000\) time steps to allow the system to stabilize. The mean and maximum polarization were then computed over the subsequent \(200\) time steps.}
        \label{Mathematical model_3D plots}
    \end{subfigure}
    \caption{Generalized cohesion model with multiple fish: polarization of the school. Parameters in both panels: \(N=100, \epsilon=40, \sigma=20, \alpha=1.5, \mu=0.2, c_p=0.002, R=100, \tau=1\).}
    \label{fig:cohesion_polarization}
\end{figure}

We can see that the school is able to maintain high level of polarization, thus a parallel behavior through a long time (Figure \ref{Polarization_mathematical_1}). 

Next we explore how the weight parameters (\(c_w, c_f\)) affect the maximum polarization of the school. Since in the Equation \ref{mathematical_model_equation}, the resulting vector is then scaled to unit length, it is actually the ratio of \((c_w, c_f, c_p)\) that matters. Thus we keep \(c_p\) constant and vary the other two. We compare it to the generalized force-based model, which has \(c_p=0\). As we can see, both models have similar shape, with generalized cohesion model's curve shifted (Figure \ref{Mathematical model_3D plots}).  For low values of \(c_w\), both the mean and maximum polarization remain close to zero, indicating swarming behavior. In contrast, for high values of \(c_w\) and relatively low values of \(c_f\), both the mean and maximum polarization are close to \(1\). This corresponds to parallel behavior, in which the school moves in a common direction with a high degree of alignment. The similarity between the mean and maximum polarization indicates the system's ability to sustain a consistently high level of polarization.

\begin{figure}[H]
    \centering
    \begin{subfigure}[t]{0.48\textwidth}
        \centering
        \includegraphics[width=1\linewidth,height=0.28\textheight,keepaspectratio]{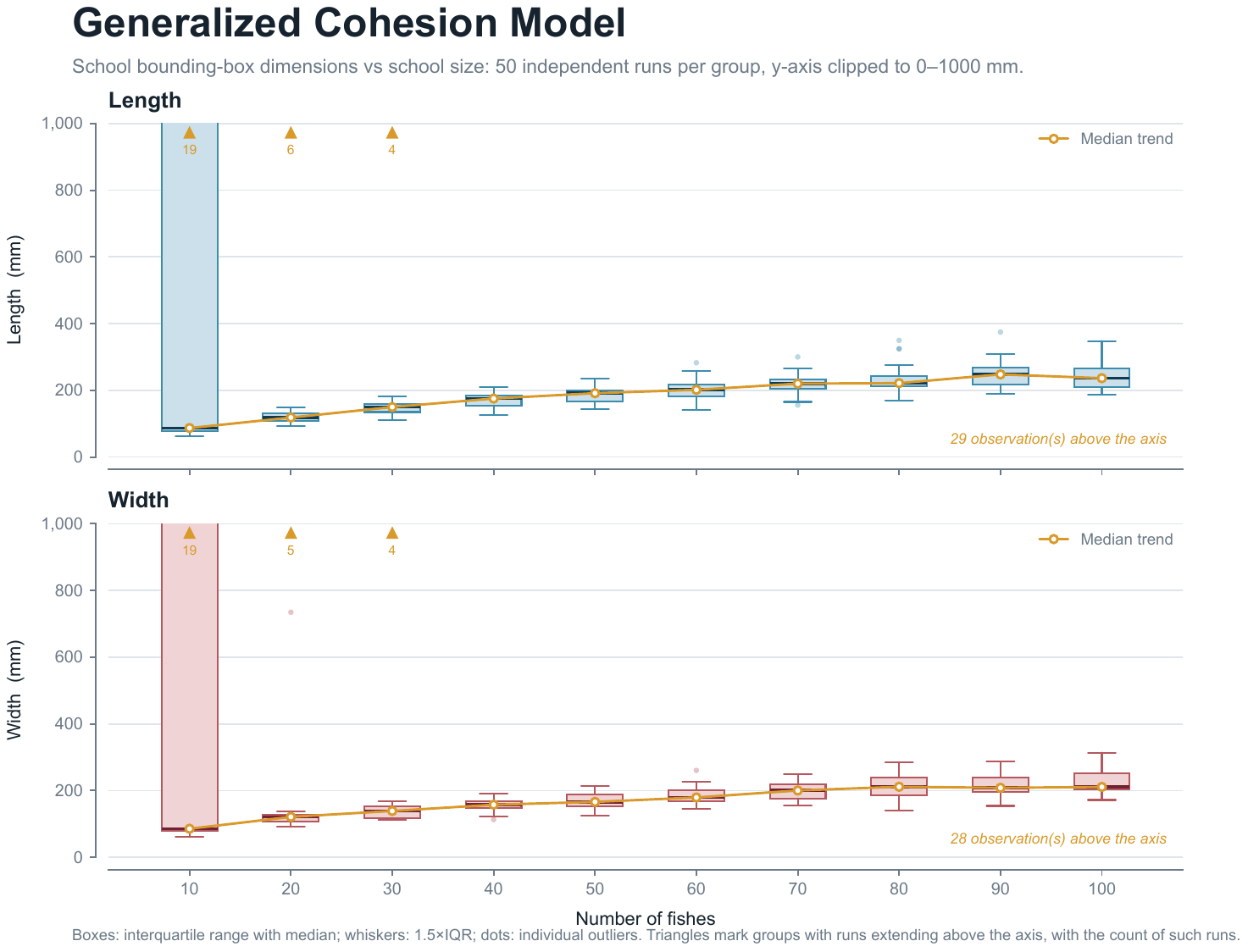}
        \caption{\footnotesize Length and width of the school vs number of fish \((N)\). Each boxplot illustrates 50 examples. In each example, the school was initially run for \(5000\) time steps, then the mean length/width was taken from the subsequent \(100\) time steps.}
        \label{Mathematical model_length width vs N}
    \end{subfigure}
    \hfill
    \begin{subfigure}[t]{0.48\textwidth}
        \centering
        \includegraphics[width=1\linewidth,height=0.28\textheight,keepaspectratio]{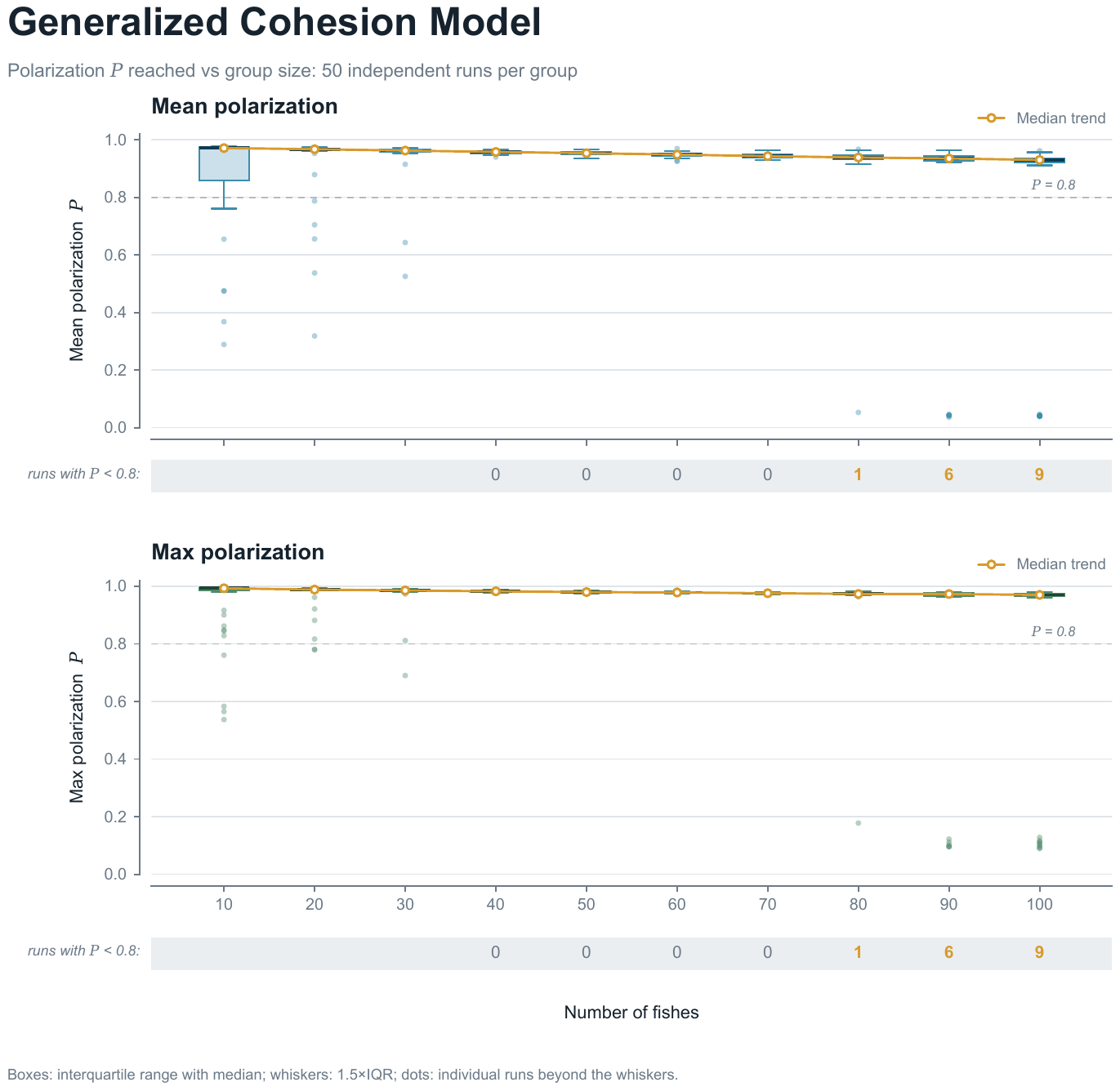}
        \caption{\footnotesize Polarization of the school vs number of fish \((N)\). Each boxplot illustrates 50 examples. In each example, the school was initially run for \(5000\) time steps, then the mean/max polarization was taken from the subsequent \(100\) time steps.}
        \label{Mathematical model_polarization vs N}
    \end{subfigure}

    \vspace{0.4cm}

    \begin{subfigure}{\textwidth}
        \centering
        \includegraphics[width=1.5\textwidth,height=0.32\textheight,keepaspectratio]{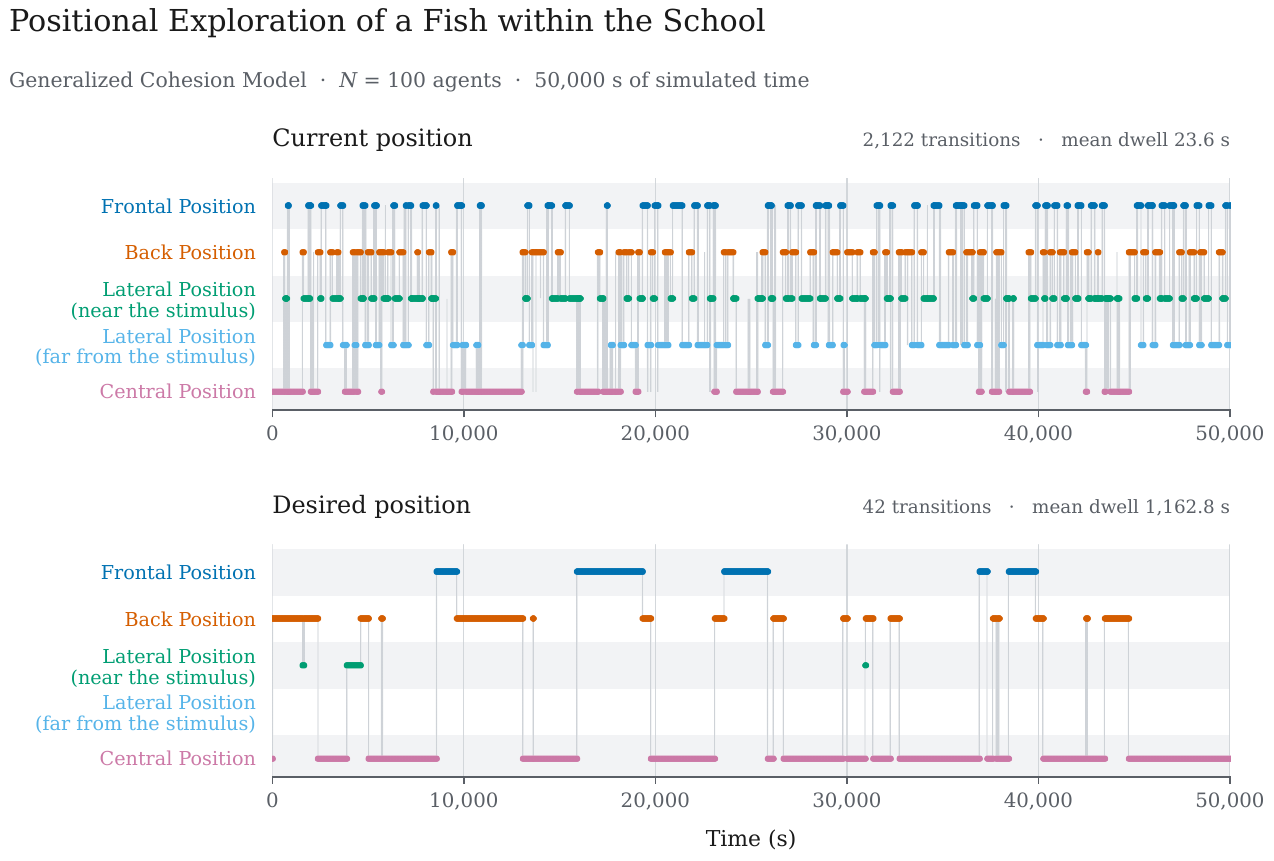}
        \caption{\footnotesize The current and desired position of the first fish, through time, with \(N=100\) fish.}
        \label{Mathematical model_position}
    \end{subfigure}
    \caption{Generalized cohesion model with multiple fish: school structure and individual exploration. Parameters in all panels: \(\epsilon=40, \sigma=20, \alpha=1.5, \mu=0.2, c_w=0.8, c_f=0.4, c_p=0.002, R=100, \tau=1\).}
    \label{fig:cohesion_school}
\end{figure}

Next, we investigate the dependence of the school's dimensions from the size (Figure \ref{Mathematical model_length width vs N}). Large values of the length and width indicate that the fish failed to merge into a single school, whereas values that remain small after \(5000 \text{ s}\) indicate that a single school has formed. As we can see, the extremely high outliers of width and length suggest that the fish sometimes struggle to form a single school when \(N \leq 30\). However, the fish formed a single school in \(100 \%\) of cases when \(N \geq 40\).

We next investigate mean and max polarization with respect to different sizes of school. We observed mainly near \(1\) polarization values for \(N \leq 30\), along with several moderate polarization values (Figure \ref{Mathematical model_polarization vs N}). These moderate values are likely due to the formation of multiple schools, as observed earlier for \(N \leq 30\). For \(40 \leq N \leq 70\), we observed exclusively \textit{parallel} behavior. For larger school sizes, however, \textit{swarming} behavior may occur, as indicated by polarization values of approximately \(0.1\).

To investigate the exploration behavior of a single fish within the school, we tracked its actual and desired positions over \(50000\) seconds from the initialization of the school (Figure~\ref{Mathematical model_position}). As the desired position is updated
only when the fish reaches it, the frequent changes in the desired position shown in the lower plot indicate that, regardless of which position the fish selects, it eventually reaches it. Additionally, the upper plot indicates extensive exploration over the observed period.

\section{Discussion}
\label{discussion}

We have evaluated several classical models (two versions of a generalized force based model, and one range based model) across \(10^5\) parameter choices. We observed, that all of them, with a high precision choice of parameters, were able to quite closely replicate the experimental observations in the no-stimuli context. The Figures \ref{fig:lopez-const}, \ref{fig:lopez-lj}, \ref{fig:couzin1} illustrate specific combinations of parameters, which successfully mimic the experimental distributions.
The generalized force-based model, when extended to predator context, was again able to replicate the experimental observations when parameters were chosen precisely. The Figure \ref{Lopez model_predator_performance_error} illustrates a specific combination of parameters which successfully mimics the experimental distribution.

However, we observed that the classical models are very sensitive to parameter change, as only less than \(0.2\%\) of the runs fell below the tolerance, though quite wise choice of ranges of parameters. The violin plots in the Figure \ref{ABC_error_total} a-c indicate that almost all the runs reproduced an error far above the baseline and tolerance. In addition, by investigating the closest parameter combinations to a specific successful parameter combination, we observed that the agent fails to match the experimental results when a minor perturbation in parameters occurs. These results suggest that classical models are not robust to represent the behaviour of the fish, as nature is usually robust to minor perturbations.

To cope with that issue, we proposed a new model (generalized cohesion model), which extends the generalized force-based model by introducing a position force, which can be also viewed as a generalized cohesion force. All the parameters other than the additional position force weight were chosen from the same ranges as the generalized force-based model during the experiments. More than \(10\%\) of the runs fell below the tolerance (Figure \ref{Mathematical model_ABC_error_total}). For a specific successful parameter combination, we observed that it is robust towards small errors in parameters (Table~\ref{tab:knn_mathematical}).

We additionally evaluated the generalized cohesion model in the multiple fish environment. First we wanted to ensure that the new model was still able to reproduce schooling behaviour and common schooling patterns. For a specific fixed parameter set, we observed (Figure \ref{Mathematical model_length width vs N}) that for adequately large school size (\(N \geq 40\)), fish were able to form a single school and do not split during a large period of time (\(5000\) sec). They sometimes struggle for smaller sizes, but this simply happens because the fish are initialised uniformly within a circle of radius \(150 \text{mm}\); with a neighbourhood radius of \(R=100\), two subgroups can be separated by more than \(R\), so that no individual of one perceives any member of the other. For sufficiently small size \((N \leq 70)\), we observed \textit{parallel} behaviour, while for larger sizes we observed a bifurcation: same set of parameters, which mainly produced a \textit{parallel} behaviour, had a chance of producing a \textit{swarming} behaviour (Figure \ref{Mathematical model_polarization vs N}). For a fixed school size of \(N=100\), by varying the weight parameters, the fish were able to reproduce the swarming and parallel behaviours (Figure \ref{Mathematical model_3D plots}), just like other models, including the generalized force-based model. In addition, the school was able to maintain a highly polarized, parallel movement through vast amount of time. The Figure \ref{Polarization_mathematical_1} shows a polarization steadily near \(1\). The average direction varies substantially over the same period, which we interpret as exploratory reorientation of the school as a whole. To additionally investigate the explorative behaviour of a single fish inside the school, we tracked its current and desired position through a long period of \(50000\) sec (Figure \ref{Mathematical model_position})). The fish changes its desired position only when it is able to reach it and remain there for \(\Delta t\) seconds. The persistent change of the desired position suggests that the fish is eventually able to reach the position it prefers. The stable and frequent change of the current position shows that the fish is constantly exploring the space around it.

From an ethological perspective, these results support a view of schooling in which an individual's position is not merely a passive geometrical consequence of attraction, repulsion and alignment. Positions within a fish school differ in their potential costs and benefits. Peripheral and frontal individuals may encounter resources or environmental information earlier, but they can also experience greater exposure to predators, whereas more central positions may provide stronger dilution and concealment benefits while increasing competition with nearby conspecifics \cite{krause2002living, ward2016sociality}. A tendency to occupy or explore particular positions may therefore represent a context-dependent behavioural decision through which an individual balances safety, access to information and access to resources.

In this interpretation, the position force should not necessarily be understood as an additional physical force directly perceived by the fish. Instead, it provides a mathematical representation of a latent behavioural tendency to seek a position relative to the group. Such a tendency may integrate several biological processes that are not explicitly separated in the present model, including individual risk sensitivity, motivational state, previous social experience and the perception of external cues. This interpretation is consistent with the increasing recognition that individuals within animal groups are not interchangeable and that behavioural heterogeneity can affect cohesion, leadership, information transfer and collective performance \cite{jolles2020heterogeneity}.

The contrast between the classical models and the generalized cohesion model is also relevant from an evolutionary perspective. Natural selection does not act directly on global quantities such as polarization or school shape, but on individuals whose behavioural responses influence survival and reproductive success. Similar collective patterns may consequently arise from different combinations of individual rules. A biologically informative model should therefore reproduce not only a particular collective configuration, but also the flexible positional behaviour from which that configuration emerges. The larger region of parameter space compatible with the observations in the generalized cohesion model suggests that the inclusion of positional preference may provide a more tolerant mechanism for generating collective organisation, without requiring a narrowly tuned balance among attraction, repulsion and alignment.

This result also illustrates a central property of natural complex systems: collective robustness can coexist with continuous variability at the individual level. The simulated fish repeatedly changed their desired and occupied positions while the school retained cohesion and, under appropriate parameter combinations, sustained highly polarized motion. Individual exploration did not necessarily destabilize the collective pattern; instead, it was accommodated within it. This suggests that exploration and cohesion need not be competing properties. Flexible movement among spatial roles may contribute to a school's capacity to acquire information and respond to changing ecological conditions while preserving collective organisation \cite{couzin2009collective, sumpter2008information, berdahl2013emergent}.

The predator condition further emphasizes the context dependence of individual behaviour. Predation is a major selective pressure shaping the evolution of fish grouping, and experimental work has shown that predators can select for coordinated collective motion \cite{ioannou2012predatory}. In a natural school, however, avoidance emerges through reciprocal interactions: the response of one fish alters the sensory environment and subsequent decisions of its neighbours. The fish--robot system is valuable because the robotic school remains unresponsive, allowing the focal individual's contribution to be isolated from this feedback. At the same time, this experimental control means that the present results describe how an individual responds to a standardized social configuration, rather than the complete feedback dynamics of a fully biological school.

The transition from a single focal fish to multiple simulated individuals should therefore be interpreted as a mechanistic hypothesis about how empirically observed positional tendencies might scale up to collective behaviour. The emergence of cohesion, swarming and polarized motion shows that these tendencies are compatible with canonical collective states. It does not yet demonstrate that every fish applies the same rule or that positional decisions are independent in a natural school. In biological groups, individuals may differ consistently in their preferred positions, and their decisions may depend on the identities, movements and previous choices of their neighbours. Incorporating these sources of heterogeneity will be important for assessing how individual variation affects the stability and adaptability of the resulting collective dynamics.








\section{Conclusions}
\label{conclusions}

This study examined whether collective schooling patterns can be reconstructed from the behaviour of an individual fish interacting with a controlled robotic school. 
To this aim, a single fish-- multi-robot system has been considered \cite{romano2021individual} that allows isolating the behavior of the single fish during schooling and in the presence of external stimuli, because the robotic school remains unresponsive and does not provide any feedback. This experimental control means that the present results describe how an individual responds to a standardized social configuration. To the best of our knowledge, both the experimental set-up and the modeling remain unique in the literature. 

We were able to reply to the initial research questions: 1) Classical force-based and range-based models reproduced the experimental distributions only within restricted regions of parameter space, whereas the generalized cohesion model remained compatible with the observations across a substantially broader range of parameter combinations. 2) Introducing a positional tendency therefore increased robustness while preserving cohesion, swarming, and highly polarized collective motion of multiple agents. 3) In presence of external stimuli, like a predator, classical models were able to closely replicate the experimental distributions, but only in very restricted regions of parameter space. An extension of the generalized cohesion model to the predator case will be investigated in the next study. 

Biologically, the results suggest that an individual's position within a school should not be treated solely as a passive consequence of attraction, repulsion and alignment. Positional choices may integrate trade-offs among predation risk, access to resources and social information. When extended to multiple agents, these individual tendencies generated stable collective organisation while allowing continued spatial exploration, illustrating how behavioural variability and group-level coherence can coexist in natural complex systems. 

In future work, we want to explore further aspects of the individual role in collective motion to overcome some of the current limitations.  
i) The classical models here tested regard only some type of force-based models, and one range-based model. In the future, we may replace the functions \(f, w\) with neural networks to represent any function, thus generalizing our result to broader type of models.
ii) In the construction of the generalized cohesion model, we assumed no correlation between the positions. In the future, we would like to empirically obtain the transition matrix between these position. In addition, we would like to take videos of the experiments, to visually assess how realistic do the modeled agents behave in the simulations. 
The current model, in fact, is based on a mechanistic hypothesis rather than a complete representation of biological schooling. Successive positional choices, residence times, and individual heterogeneity must be quantified from trajectory data, and reciprocal interactions should be tested in fully responsive groups. Combining controlled fish--robot experiments with these measurements could clarify how context-dependent individual decisions scale into adaptive collective behaviour.






\section*{Acknowledgments} 

V. Grigoryan wrote all simulation codes, run the numerical experiments, analyzed the results, and wrote the first draft of the paper. D. Romano and C. Stefanini conceived the initial experiment and supervised the outcome and interpretation of the results. G. De Masi conceived and supervised all the modeling part, with overarching supervision on the study. D.R., C.S., G.D.M. finilized the paper. This study is part of the Project ``Artificial intelligence and statistical models to understand the behavior of social marine species”, 2025-2026, funded by SAFIR, Sorbonne University Abu Dhabi. Disclaimer: AI tools were used to improve English grammar and readability of the paper. Data will be provided upon request to guarantee reproducibility. The authors declare that they have no competing interests.

\clearpage
\section*{Supplementary Materials}
\label{SupplementaryMaterials}
\setcounter{figure}{0}
\setcounter{table}{0}
\renewcommand{\thefigure}{S\arabic{figure}}
\renewcommand{\thetable}{S\arabic{table}}

\noindent Fig. S1. The distribution of the error across the \(100000\) parameter combinations: generalized force-based model with constant alignment, predator context.

\noindent Fig. S2. Local sensitivity of the generalized force-based model with constant alignment in the predator context, around run 11442.

\noindent Table S1. Generalized force-based model with constant alignment: parameter combinations whose error falls below the tolerance and have a non-zero frontal position mean.

\noindent Table S2. Generalized force-based model with positive Lennard--Jones alignment: parameter combinations whose error falls below the tolerance and have a non-zero frontal position mean.

\noindent Table S3. The range-based model: parameter combinations whose error falls below the tolerance and have a non-zero frontal position mean.

\begin{figure}[H]
    \centering
    \includegraphics[width=0.5\textwidth]{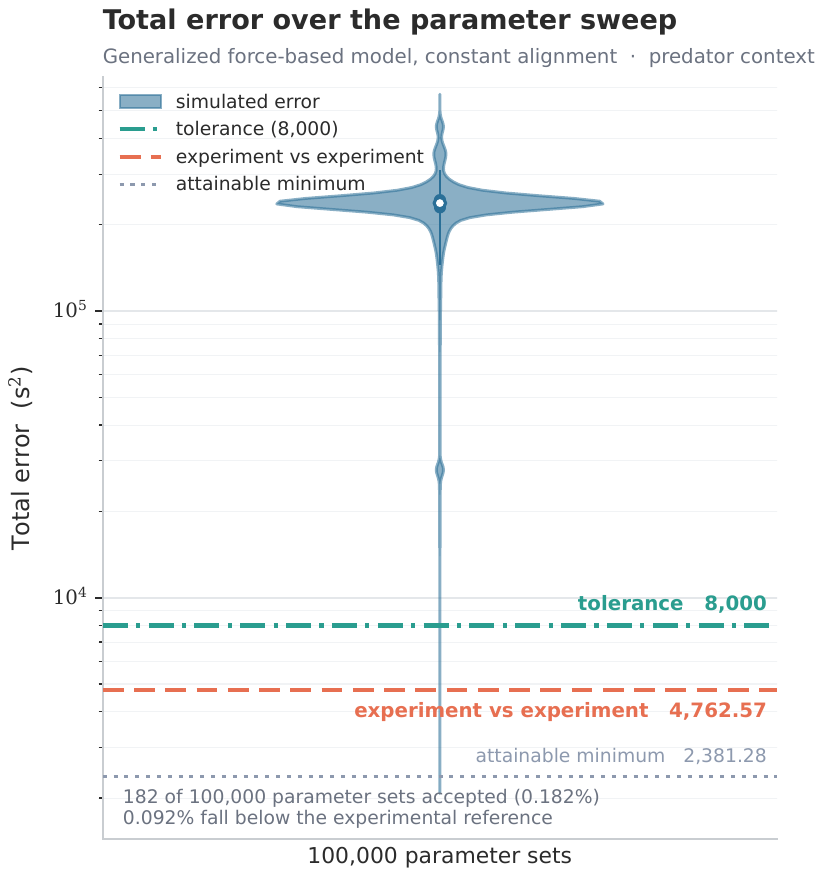}
    \caption{\footnotesize The distribution of the error across the \(100000\) parameter combinations: generalized force-based model with constant alignment, predator context.}
    \label{Lopez model_predator_ABC_error_total}
\end{figure}

\begin{figure}[H]
\centering
\begin{subfigure}{\textwidth}
\centering
\begin{adjustbox}{max width=\textwidth}
\footnotesize
\begin{tabular}{lrrrrrrrrrr}
\toprule
Run ID & $\sigma$ & $\alpha$ & $c_f$ & $s$ & $\mu$ & $c_P$ & $r_0$ & $\beta$ & $D$ & Error \\
\midrule
\textbf{11442} & \textbf{17.34} & \textbf{2.60} & \textbf{6132.64} & \textbf{17.24} & \textbf{0.27} & \textbf{38.71} & \textbf{311.24} & \textbf{0.082} & \textbf{0} & \textbf{2866.13} \\
10863 & 19.35 & 2.76 & 3658.70 & 16.08 & 0.45 &  71.62 & 306.18 & 0.083 & 0.239 &  51367.70 \\
8771  & 16.39 & 2.42 & 4521.45 & 17.87 & 0.34 & 149.68 & 360.83 & 0.092 & 0.241 & 238319.14 \\
57910 & 19.71 & 2.03 & 4930.50 & 17.59 & 0.33 & 180.80 & 306.12 & 0.072 & 0.242 & 202245.77 \\
875   & 18.68 & 2.64 & 9162.44 & 19.48 & 0.42 & 103.49 & 316.38 & 0.076 & 0.247 &  22043.24 \\
82590 & 17.48 & 3.51 & 4736.53 & 17.21 & 0.22 &   8.73 & 337.92 & 0.092 & 0.257 & 224285.64 \\
\midrule
\textbf{Experiment} & --- & --- & --- & --- & --- & --- & --- & --- & --- & \textbf{4762.57} \\
\bottomrule
\end{tabular}
\end{adjustbox}
\caption{Reference combination (bold) and its five nearest neighbours,
ordered by distance $D$. $\epsilon=1$ and $c_w=1$ in all rows.}
\label{tab:knn_lopez_predator}
\end{subfigure}

\vspace{0.4cm}
\begin{subfigure}{0.55\textwidth}
\centering
\includegraphics[width=\textwidth]{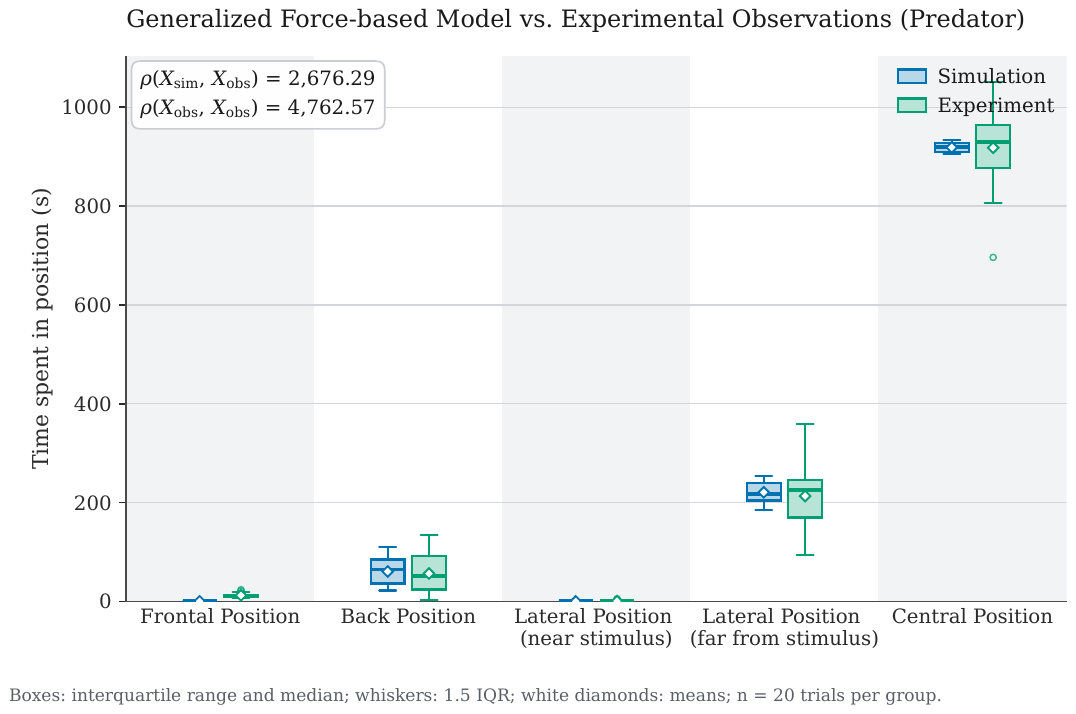}
\caption{Performance of the reference combination (run 11442, different seed).}
\label{Lopez model_predator_performance_error}
\end{subfigure}
\caption{Local sensitivity of the generalized force-based model with constant
alignment in the predator context, around run 11442.}
\label{fig:knn_lopez_predator}
\end{figure}

\begin{table}[H]
\centering
\caption{Generalized force-based model with constant alignment: parameter
combinations whose error falls below the tolerance and have a non-zero frontal
position mean. The final row gives the experimental reference values.}
\label{tab:lopez-const}
\begin{adjustbox}{max width=\textwidth}
\footnotesize
\begin{tabular}{llrrrrrrrrr}
\toprule
Case & Run ID & $\epsilon$ & $\sigma$ & $\alpha$ & $c_w$ & $c_f$ & $s$ &
$\mu$ & Error & FP mean \\
\midrule
1 & 19179 & 1 & 15.80 & 2.37 & 1 &  58.49 &  9.63 & 0.26 & 7612.32 &   6.73 \\
2 & 31384 & 1 &  9.87 & 1.52 & 1 &  94.06 &  9.13 & 0.27 & 8685.55 &   1.43 \\
3 & 32856 & 1 & 11.83 & 1.05 & 1 & 123.58 & 15.04 & 0.84 & 8884.52 &   7.54 \\
4 & 44303 & 1 &  9.12 & 1.13 & 1 &  91.56 & 11.44 & 0.58 & 8831.50 &   5.65 \\
5 & 66481 & 1 &  2.35 & 1.32 & 1 & 156.90 &  9.76 & 0.45 & 9239.12 &  33.48 \\
6 & 67117 & 1 &  1.53 & 1.68 & 1 & 481.63 &  8.45 & 0.10 & 5125.49 &  63.42 \\
7 & 71143 & 1 &  1.21 & 1.03 & 1 & 151.63 &  9.59 & 0.44 & 7445.44 &  61.96 \\
8 & 75580 & 1 & 13.95 & 1.11 & 1 &  31.82 &  9.62 & 0.03 & 9622.08 & 194.91 \\
9 & 91490 & 1 & 11.18 & 2.17 & 1 & 252.50 & 19.56 & 0.90 & 9952.97 &   0.22 \\
\midrule
\multicolumn{2}{l}{\textbf{Experiment}} & --- & --- & --- & --- & --- & --- &
--- & \textbf{7652.24} & \textbf{98.03} \\
\bottomrule
\end{tabular}
\end{adjustbox}
\end{table}

\begin{table}[H]
\centering
\caption{Generalized force-based model with positive Lennard--Jones alignment:
parameter combinations whose error falls below the tolerance and have a
non-zero frontal position mean. The final row gives the experimental reference
values.}
\label{tab:lopez-lj}
\begin{adjustbox}{max width=\textwidth}
\footnotesize
\begin{tabular}{llrrrrrrrrr}
\toprule
Case & Run ID & $\epsilon$ & $\sigma$ & $\alpha$ & $c_w$ & $c_f$ & $s$ &
$\mu$ & Error & FP mean \\
\midrule
1 &  9416 & 1 &  1.06 & 5.70 & 1 &  0.57 & 11.43 & 0.09 & 6691.32 &   2.39 \\
2 & 12049 & 1 & 13.12 & 1.71 & 1 & 26.05 & 17.10 & 0.94 & 9598.67 &   0.01 \\
3 & 82633 & 1 &  7.93 & 5.06 & 1 &  0.85 & 11.07 & 0.07 & 5726.12 & 111.71 \\
4 & 85973 & 1 & 13.10 & 1.43 & 1 &  6.98 & 11.41 & 0.52 & 9293.76 &  11.93 \\
\midrule
\multicolumn{2}{l}{\textbf{Experiment}} & --- & --- & --- & --- & --- & --- &
--- & \textbf{7652.24} & \textbf{98.03} \\
\bottomrule
\end{tabular}
\end{adjustbox}
\end{table}

\begin{table}[H]
\centering
\caption{The range-based model: parameter combinations whose error falls below
the tolerance and have a non-zero frontal position mean. The noise amplitude
$\sigma$ is reported in units of $10^{-2}$ so that two decimals remain
informative. The final row gives the experimental reference values.}
\label{tab:couzin}
\begin{adjustbox}{max width=\textwidth}
\footnotesize
\begin{tabular}{llrrrrrrrrr}
\toprule
Case & Run ID & $r_r$ & $\Delta r_o$ & $\Delta r_a$ & $\alpha$ & $\theta$ & $s$ &
$\sigma\,(\times 10^{-2})$ & Error & FP mean \\
\midrule
 1 &  2606 & 22.83 &  8.84 & 29.87 & 324.83 &  86.31 & 11.92 &  6.89 & 7014.83 &  87.48 \\
 2 &  6900 &  7.97 & 39.43 & 35.04 & 272.07 &  23.52 &  8.98 & 13.21 & 7656.62 &   1.26 \\
 3 & 10153 & 21.28 & 10.85 & 85.00 & 243.09 &  50.98 & 11.68 &  1.72 & 7301.13 & 179.89 \\
 4 & 15226 &  2.89 &  2.57 & 84.24 & 296.84 &  39.47 & 17.24 & 15.70 & 8408.05 &  14.61 \\
 5 & 17347 &  3.60 & 35.24 & 15.00 & 305.67 &  36.48 &  9.24 & 10.57 & 9209.49 &   0.46 \\
 6 & 19634 &  2.26 &  8.79 & 84.47 & 291.75 &  40.60 & 17.09 & 10.81 & 9604.81 &   6.59 \\
 7 & 26578 & 22.84 &  5.41 & 62.13 & 328.69 &  38.22 & 11.71 & 19.59 & 9356.86 &  85.16 \\
 8 & 30285 & 15.00 & 20.90 & 68.56 & 236.75 & 106.23 &  9.54 &  3.87 & 9415.50 &  15.31 \\
 9 & 41761 &  2.00 & 48.13 & 18.41 & 211.76 &  86.96 &  8.65 & 14.58 & 7800.93 &   0.43 \\
10 & 44356 &  6.51 & 44.28 & 29.76 & 243.96 &  85.37 &  8.87 & 17.16 & 9926.87 &   8.36 \\
11 & 44642 &  8.60 &  1.79 & 30.33 & 283.02 & 166.14 & 10.69 &  1.42 & 9612.84 &   2.11 \\
12 & 48575 &  9.26 & 38.14 & 19.26 & 296.80 &  50.71 &  8.77 & 10.41 & 7969.07 &   1.25 \\
13 & 52206 &  5.52 & 44.35 & 67.50 & 205.78 & 110.59 &  8.97 & 18.78 & 8319.36 &  16.33 \\
14 & 52440 &  4.02 & 35.55 & 13.13 & 344.02 & 157.88 &  9.17 &  8.96 & 8603.85 &   0.69 \\
15 & 55449 &  3.45 & 35.95 & 35.91 & 212.18 &  20.07 &  9.16 & 19.08 & 9743.50 &   1.07 \\
16 & 57100 &  1.12 &  8.26 & 63.30 & 271.08 &  46.47 & 18.48 & 14.85 & 8394.34 &  13.43 \\
17 & 58624 &  9.55 &  1.28 & 68.40 & 203.52 &  12.79 & 10.33 & 13.94 & 9183.70 &   8.33 \\
18 & 59340 &  9.55 & 18.52 & 24.96 & 290.44 & 143.28 &  9.56 &  9.63 & 7747.54 &   0.29 \\
19 & 66171 & 21.77 &  2.93 & 82.25 & 354.13 &  33.26 & 11.99 &  4.77 & 8388.29 &  16.07 \\
20 & 66849 &  3.15 & 27.65 & 20.92 & 211.66 &  29.90 &  9.66 &  0.35 & 9596.27 &   0.25 \\
21 & 68133 & 13.12 & 19.18 & 24.57 & 306.02 &  56.88 &  9.48 & 17.99 & 9006.78 &   0.60 \\
22 & 68354 & 16.23 & 13.46 & 56.30 & 241.96 & 177.63 &  9.75 & 14.91 & 9849.17 &   7.69 \\
23 & 79441 & 14.50 & 13.30 & 20.26 & 292.92 & 147.97 &  9.71 & 18.13 & 7204.41 &   1.26 \\
24 & 82706 & 18.66 & 14.21 & 50.06 & 209.29 & 149.10 &  9.62 & 18.63 & 9873.37 &   2.13 \\
25 & 91571 & 10.40 &  2.44 & 28.29 & 359.96 & 101.98 & 14.70 & 13.46 & 9480.14 & 157.46 \\
26 & 92818 & 20.63 & 13.91 & 83.45 & 261.65 & 113.74 &  9.74 & 12.57 & 8346.75 &   0.69 \\
27 & 99712 & 19.33 & 22.55 & 54.36 & 334.53 &  27.07 &  9.47 & 16.23 & 8036.37 &   0.77 \\
\midrule
\multicolumn{2}{l}{\textbf{Experiment}} & --- & --- & --- & --- & --- & --- &
--- & \textbf{7652.24} & \textbf{98.03} \\
\bottomrule
\end{tabular}
\end{adjustbox}
\end{table}

\printbibliography

\end{document}